\documentclass[fleqn,usenatbib]{mnras}
    
\usepackage{newtxtext,newtxmath}
\usepackage[T1]{fontenc}
\usepackage{ae,aecompl}

\usepackage{graphicx}	 
\usepackage{amsmath}	
\newcommand{\hii}{H$_2\,$}	

\usepackage{color}
\usepackage{gensymb}

\usepackage{scalerel,tikz}
\usetikzlibrary{svg.path}
\definecolor{orcidlogocol}{HTML}{A6CE39}
\tikzset{orcidlogo/.pic={
 \fill[orcidlogocol] svg{M256,128c0,70.7-57.3,128-128,128C57.3,256,0,198.7,0,128C0,57.3,57.3,0,128,0C198.7,0,256,57.3,256,128z};
 \fill[white] svg{M86.3,186.2H70.9V79.1h15.4v48.4V186.2z}
 svg{M108.9,79.1h41.6c39.6,0,57,28.3,57,53.6c0,27.5-21.5,53.6-56.8,53.6h-41.8V79.1z M124.3,172.4h24.5c34.9,0,42.9-26.5,42.9-39.7c0-21.5-13.7-39.7-43.7-39.7h-23.7V172.4z}
 svg{M88.7,56.8c0,5.5-4.5,10.1-10.1,10.1c-5.6,0-10.1-4.6-10.1-10.1c0-5.6,4.5-10.1,10.1-10.1C84.2,46.7,88.7,51.3,88.7,56.8z};
}}
\newcommand\orcidicon[1]{\href{https://orcid.org/#1}{\mbox{\scalerel*{
\begin{tikzpicture}[yscale=-1,transform shape]
\pic{orcidlogo};
\end{tikzpicture}
}{|}}}}

\newcommand{\aref}[1]{\hyperref[#1]{Appendix~\ref{#1}}}
\defcitealias{2020arXiv200211502S}{SFK20}

\usepackage[normalem]{ulem}

\definecolor{darkgreen}{rgb}{0.13, 0.55, 0.13}
\definecolor{brown}{rgb}{0.65, 0.16, 0.16}

\title[Dynamo amplification in first stars]{Magnetic field amplification in accretion discs around the first stars: implications for the primordial IMF}

\author[P. Sharda et al.]{
Piyush Sharda$^{\orcidicon{0000-0003-3347-7094}\,1,2}$\thanks{piyush.sharda@anu.edu.au (PS)},
Christoph Federrath$^{\orcidicon{0000-0002-0706-2306}\,1,2}$\thanks{christoph.federrath@anu.edu.au (CF)},
Mark R. Krumholz$^{\orcidicon{0000-0003-3893-854X}\,1,2}$\thanks{mark.krumholz@anu.edu.au (MRK)}, and
\newauthor
Dominik R. G. Schleicher$^{\orcidicon{0000-0002-9642-120X}\,3}$
\\
$^{1}$Research School of Astronomy and Astrophysics, Australian National University, Canberra, ACT~2611, Australia\\
$^{2}$Australian Research Council Centre of Excellence for All Sky Astrophysics in 3 Dimensions (ASTRO 3D), Australia\\
$^{3}$Departamento de Astronom\'ia, Facultad Ciencias F\'isicas y Matem\'aticas, Universidad de Concepci\'on, Casilla 160-C, Concepci\'on, Chile\\
}

\date{Accepted 2021 February 22. Received 2021 February 17; in original form 2020 July 6}
    
\pubyear{2021}

\begin{document}
\label{firstpage}
\pagerange{\pageref{firstpage}--\pageref{lastpage}}
\maketitle

\begin{abstract}
Magnetic fields play an important role in the dynamics of present-day molecular clouds. Recent work has shown that magnetic fields are equally important for primordial clouds, which form the first stars in the Universe. While the primordial magnetic field strength on cosmic scales is largely unconstrained, theoretical models strongly suggest that a weak seed field existed in the early Universe. We study how the amplification of such a weak field can influence the evolution of accretion discs around first stars, and thus affect the primordial initial mass function (IMF). We perform a suite of 3D ideal magneto-hydrodynamic (MHD) simulations with different initial field strengths and numerical resolutions. We find that, in simulations with sufficient spatial resolution to resolve the Jeans scale during the collapse, even initially weak magnetic fields grow exponentially to become dynamically important due to both the so-called \emph{small-scale turbulent dynamo} and the \emph{large-scale mean-field dynamo}. Capturing the small-scale dynamo action depends primarily on how well we resolve the Jeans length, while capturing the large-scale dynamo depends on the Jeans resolution as well as the maximum absolute resolution. Provided enough resolution, we find that fragmentation does not depend strongly on the initial field strength, because even weak fields grow to become strong. However, fragmentation in runs with magnetic fields differs significantly from those without magnetic fields. We conclude that the development of dynamically strong magnetic fields during the formation of the first stars is likely inevitable, and that these fields had a significant impact on the primordial IMF.
\end{abstract}


\section{Introduction}
\label{s:intro}
From the formation of molecular clouds to their collapse into protostar-accretion disc systems, turbulence and magnetic fields play several roles in setting the overall direction for a star formation episode. While extensive studies have been carried out to investigate the role of turbulent magnetic fields in present-day star formation (see reviews by \citealt{2012ARA&A..50...29C,2017ARA&A..55..111H,2018FrASS...5...39W,2019FrASS...6....5H,2019FrASS...6....7K,2019FrASS...6...66C,2020SSRv..216...43Z}), only a handful of 3D simulations have looked at their role in the early Universe, especially during the formation of the first generation of stars \citep{2008ApJ...685..690M,2010ApJ...721L.134S,2010A&A...522A.115S,2012ApJ...745..154T,2013MNRAS.432..668L,2013MNRAS.435.3283M,2014MNRAS.440.1551L,2019arXiv191107898L,2019MNRAS.487.4525G}. This is primarily due to the lack of solid constraints on the magnetic field strength and topology in the early Universe \citep{2002RvMP...74..775W,2004IJMPD..13..391G,2012SSRv..166...37W,2012SSRv..166....1R,2014PhRvD..89j3001W,2016RPPh...79g6901S}. However, there is a growing consensus on the presence of a cosmic-scale primordial field, no matter how weak \citep{1996PhRvD..54.1291B,2012arXiv1209.1438H,2016RPPh...79g6901S,2016A&A...594A..19P}. This motivates studying magnetic fields that may be amplified from the primordial field during the collapse of molecular clouds, leading to Population III star formation. 

Several studies have conclusively shown that the presence of a turbulent dynamo \citep{1968JETP...26.1031K,1981PhRvL..47.1060M,2005PhR...417....1B,2016RPPh...79g6901S} can exponentially amplify any weak seed field to near-saturation values (e.g., \citealt{2011ApJ...731...62F,2012ApJ...754...99S,2014ApJ...797L..19F,2015PhRvE..92b3010S,2016JPlPh..82f5301F,2016ApJ...833..215X,2020arXiv200614607M}). In the early Universe, the presence of such a turbulent dynamo driven by gravity is expected when baryonic matter starts collapsing towards the centre of dark matter minihaloes \citep{2008MNRAS.387.1021G,2008ApJ...682..745W,2012ApJ...745..154T,2019MNRAS.487.4525G}. This infall leads to the creation of overdense regions that harbour the first molecular clouds where Population III star formation ultimately takes place (see reviews by \citealt{2013RPPh...76k2901B,2019ffbh.book...67K,2020SSRv..216...48H}). Apart from the action of the small-scale turbulent dynamo, it is also expected that accretion discs around Population III stars may contain a large-scale mean field component \citep{2019arXiv191107898L}. This can occur if discs undergo differential rotation and angular momentum transport through viscous stresses, thereby generating a large-scale dynamo from a seed field that can sustain a dynamically strong and coherent mean field component \citep{1988ASSL..133.....R,1995ApJ...446..741B,1996ApJ...464..690H,1996ApJ...463..656S}. In fact, given that the characteristic diffusion timescale in accretion discs is very short ($10^2-10^4\,\mathrm{s}$) as compared to viscous timescales (order of few $\mathrm{yr}$), dynamically strong magnetic fields that last for the lifetime of the disc can only be generated by a dynamo operating in accretion discs \citep{1995A&A...298..934R}.

The expectation that dynamically-significant magnetic fields might be present during the formation of Population III stars naturally raises the question of how such fields might affect the initial mass function (IMF) of the first stars. In a recent work, \citet[hereafter, \citetalias{2020arXiv200211502S}]{2020arXiv200211502S}, we presented the first suite of 3D magneto-hydrodynamical (MHD) simulations of Population III star formation aimed at answering this question. We showed that dynamically strong magnetic fields, if present during the formation of the first stars, suppress fragmentation in primordial clouds, thereby increasing the mean stellar mass and greatly decreasing the prevalence of low-mass Population III stars that could potentially survive to the present day. Several works that include radiative transfer \citep{2009MNRAS.398...33P,2009MNRAS.392.1363B,2012MNRAS.419.3115B}, protostellar heating feedback \citep{2016MNRAS.458..673G,FederrathKrumholzHopkins2017,MathewFederrath2020}, or both \citep{2009ApJ...703..131O,2010ApJ...713.1120K,2010ApJ...710.1343U,Myers14a,2020ApJ...904..194H} highlight that radiation feedback plays an important role in setting the present-day stellar IMF, potentially even more so than magnetic fields (\citealt{Krumholz16c,Cunningham18a,2019MNRAS.489.1719W,2019FrASS...6....7K,2019A&A...622A.125L}; see, however, \citealt{2020AJ....160...78R}, who find that, for present-day massive stars, magnetically-driven outflows are more important than radiation feedback). However, \citetalias{2020arXiv200211502S} argue that this might not be the case for Population III stars because the late onset of radiation feedback due to the absence of dust \citep{2011Sci...334.1250H,2012ApJ...760L..37H,2020ApJ...892L..14S} allows a much longer period when magnetic effects and magnetic pressure can dominate and consequently impact the primordial IMF. However, the results of \citetalias{2020arXiv200211502S} do not fully resolve the question of whether magnetic fields significantly influence the first star IMF, because they did not determine the magnetic field strength self-consistently; they only showed that, if fields near dynamo-saturation levels are present, they have a significant effect on the IMF of the first stars. Calculating the field strength self-consistently is a challenging numerical problem, because dynamo amplification is exquisitely sensitive to numerical dissipation, and thus, very high resolution is required to recover even qualitatively correct estimates for the rate of dynamo growth \citep{2014ApJ...797L..19F,2015PhRvE..92b3010S,2016JPlPh..82f5301F,2020arXiv200614607M}. The simulations of \citetalias{2020arXiv200211502S} only marginally resolve the dynamo action, and thus leave the question of the true magnetic field strength in primordial star-forming regions unsolved.

In this study, we answer this question by studying in detail how dynamo amplification can occur in first star discs. We find that, given sufficient resolution in the disc, even an initially weak field can be exponentially amplified due to the presence of both the small-scale and the large-scale dynamo; the former primarily amplifies the turbulent component of the field whereas the latter amplifies the mean component. We show that the resulting saturation level of the field is high enough that magnetic effects on the IMF are inevitably significant. The remainder of this paper is organised as follows. In \autoref{s:simulationsuite}, we describe our suite of simulations. In \autoref{s:analysis_discussions} we present our simulation results and discussions; in \autoref{s:fragmentation}, we comment on how our results can potentially impact the primordial IMF, and we summarise the implications of our findings in \autoref{s:conclusions}.

\section{Simulation suite}
\label{s:simulationsuite}
The simulations presented here are similar to those described in \citetalias{2020arXiv200211502S}, where we motivate in detail the choice of initial conditions and numerical methods. Here, we only summarise the key aspects of the simulation setup and methods. For details, we refer the reader to \citetalias{2020arXiv200211502S}.

\subsection{MHD code and basic initial conditions}
We perform 3D ideal MHD simulations of Population III star formation using the adaptive mesh refinement (AMR) code FLASH \citep{2000ApJS..131..273F,2008ASPC..385..145D}, together with the primordial chemistry network from the astro-chemistry package KROME \citep{2014MNRAS.439.2386G}. We use sink particles to represent stars \citep{2010ApJ...713..269F}; the density threshold for sink particle formation is $n_\mathrm{sink} \sim 10^{13}\,\mathrm{cm^{-3}}$. We start the simulations from a spherical core of mass $M_{\rm cl}=10^3\,\mathrm{M}_\odot$, with uniform density ($n = 9.05\times10^3\,\mathrm{cm^{-3}}$), temperature ($265\,\mathrm{K}$) and composition (with mass fractions $x_{\mathrm{H}} = 0.7502,\,x_{\mathrm{H_2}} = 0.0006,\,x_{\mathrm{He}} = 0.2492$) as appropriate for the formation of the first stars at the centre of dark matter minihaloes at a redshift of 30 (\citealt{2019MNRAS.490..513S}, and references therein). The simulation box is of size $2.4\,\mathrm{pc}$ and the boundary conditions are outflow/inflow for the hydrodynamics and isolated for computing gravitational interactions. The initial conditions also include a driven, mixed mode of turbulence \citep{2010A&A...512A..81F,2011PhRvL.107k4504F} that initially follows a velocity power spectrum $P_{\mathrm{v}} \propto k^{-1.8}$, where $k$ is the wave number that spans $2 \leq k \leq 20$. The initial Mach number is trans-sonic, such that the velocity fluctuations equal the local sound speed at the initial temperature. The maximum resolution of the simulations is $\Delta x = 7.6\,\mathrm{au}$, equivalent to a maximum effective resolution of $65,536^3$ grid cells. A limitation of our work is that we do not include radiation feedback in our simulations, which can halt accretion onto massive stars provided the accretion rates are low \citep{2011Sci...334.1250H,2012ApJ...760L..37H,2016ApJ...824..119H,2020ApJ...892L..14S}. We also discuss caveats associated with the exclusion of non-ideal MHD effects in \autoref{s:nonidealMHD}.

\subsection{Criteria for resolving dynamo action and initial conditions for the magnetic field}
In our previous simulations \citepalias{2020arXiv200211502S}, the refinement criteria were set so as to guarantee that, on all levels at or above the finest, the Jeans length \citep{2010ApJ...713..269F},
\begin{equation}
\lambda_{\mathrm{J}} = \sqrt{\frac{\pi c^2_{\mathrm{s}}}{G\rho}}
\label{eq:jeanslength}
\end{equation}
is resolved by at least 32 cells at all times (here, $c_{\mathrm{s}}$ is the sound speed). These simulations used three different initially turbulent magnetic field strengths of $1\,\mathrm{fG}$, $9\,\mathrm{\mu G}$ and $30\,\mathrm{\mu G}$. The latter two of these correspond to plausible scenarios whereby the turbulent dynamo saturates at a ratio of magnetic energy, $E_{\mathrm{mag}}$, to turbulent kinetic energy, $E_{\mathrm{turb,kin}}$, of 0.01 and 0.1, respectively \citep{2014ApJ...797L..19F,2015PhRvE..92b3010S,2016JPlPh..82f5301F}. The magnetic power spectrum goes as $P_{\mathrm{mag}} \propto k^{1.5}$ for $2 \leq k \leq 20$. We use the first and third sets, that is, runs with a field strength of $1\,\mathrm{fG}$ and $30\,\mathrm{\mu G}$, in this analysis\footnote{The statistical outcomes of the runs with an initial field strength of $9$ and $30\,\mathrm{\mu G}$ are very similar, so we use only the latter for simplicity.}. We call these runs weakJ32 and strongJ32, to represent that they start with a weak and strong field, respectively, and the Jeans length is refined with 32 cells at all times. \citetalias{2020arXiv200211502S} provide 50 realisations of each of these cases, which are identical in their mean properties, but differ in the random realisation of the turbulent velocity and magnetic fields. We use half of their suite (25 realisations of each magnetic field strength) in this study.

As we discuss in \autoref{s:intro}, dynamo simulations are extremely sensitive to resolution. We therefore repeat these earlier simulations, but at a higher resolution of 64 cells per Jeans length instead of 32 as used by \citetalias{2020arXiv200211502S}. We call these two sets of runs weakJ64 and strongJ64, respectively. Our motivation to go to higher Jeans resolution is to check the operation of the turbulent dynamo in the weak-field case; we expect the strong-field case not to show any small-scale dynamo action, since the initially turbulent field should be close to saturation. Note that a higher Jeans resolution does not mean that we resolve the grid to a smaller cell size; higher Jeans resolution simply implies that the grid creates more cells (of the same size) to better resolve the Jeans length. Thus, the minimum value of $\Delta x$ remains the same in runs between 32 and 64 cells per Jeans length. However, we also discuss two cases below where we increase the maximum resolution, but these are not part of our main simulation suite, because we are unable to perform a large number of such simulations due to computational expense. Indeed, increasing only the Jeans resolution requires substantially more computational time \citep{2011ApJ...731...62F}. For example, runs with 64 cells per Jeans length are up to 8 times more expensive than the respective runs with 32 cells per Jeans length. This increased cost of the simulations precludes us from performing higher-resolution runs for the entire suite of 50 simulations presented in \citetalias{2020arXiv200211502S}. However, Figure~7 of \citetalias{2020arXiv200211502S} indicates that 25 realisations constitute a large enough sample to allow us to recover the true statistics of the sink mass distribution with reasonable accuracy. In particular, even 25 realisations are sufficient to show a clear distinction between the distributions of sink particle masses produced in magnetised versus purely hydrodynamic simulations, which is the critical question for us. We summarise the full simulation set we use in this paper in \autoref{tab:tab1}.

\begin{table}
\centering
\caption{List of simulations used in this work. $B$ is the initial root-mean-square magnetic field strength. J represents the number of cells per Jeans length used, $\Delta x$ is the minimum cell size at the highest level of the AMR grid, and $N_{\mathrm{r}}$ is the number of realizations per run. All the realizations between the different runs are matched in pairs of initial random seeds for the turbulence and the magnetic field.}
\begin{tabular}{|l|c|c|c|c|r|}
\hline
ID & $B$& J & $\Delta x$ & $N_{\mathrm{r}}$ & Source\\
\hline
weakJ32 & $1\,\mathrm{fG}$ & 32 & $7.6\,\mathrm{au}$ & 25 & \citetalias{2020arXiv200211502S} \\
weakJ64 & $1\,\mathrm{fG}$ & 64 & $7.6\,\mathrm{au}$ & 25 & This Work\\
strongJ32 & $30\,\mathrm{\mu G}$ & 32 & $7.6\,\mathrm{au}$ & 25  & \citetalias{2020arXiv200211502S}\\
strongJ64 & $30\,\mathrm{\mu G}$ & 64 & $7.6\,\mathrm{au}$ & 25 & This Work\\
\hline
\end{tabular}
\label{tab:tab1}
\end{table}

\section{Results and Discussions}
\label{s:analysis_discussions}
Following \citetalias{2020arXiv200211502S}, we stop the runs at a time when the sink particle has accreted $50\,\mathrm{M}_{\odot}$, corresponding to a parameterized star formation efficiency, $\mathrm{SFE} = \sum M_{\mathrm{sink}}/M_{\mathrm{cl}} = 0.05$, where $M_{\mathrm{cl}} = 10^3\,\mathrm{M}_{\odot}$ is the initial cloud mass. We stop the simulations based on this criterion, because we do not include radiation feedback, which starts to play a dominant role for massive first stars \citep{2011Sci...334.1250H,2012ApJ...760L..37H,2016ApJ...824..119H,2020ApJ...892L..14S}. Note that for all the analysis except for the effects of Jeans resolution on fragmentation, we only use the subset of simulations that forms a single sink particle of mass $50\,\rm{M_{\odot}}$ ($\sim$ 8 out of the 25 realizations in each case). This is because such simulations have a well-defined accretion disc, enabling a cleaner study of the effects of the magnetic-field amplification in the disc. The simulations where secondary fragmentation takes place form more complex disc-like structures characterised by strong spiral density waves and circum-binary or circum-ternary discs. In such cases, studying the amplification of the small- and/or large-scale dynamo is challenging as it would demand that all the accretion discs be well resolved, and the full simulation be followed to a significantly longer time. Thus, we only study magnetic field amplification in accretion discs around massive first stars.

\begin{figure*}
\includegraphics[width=\columnwidth]{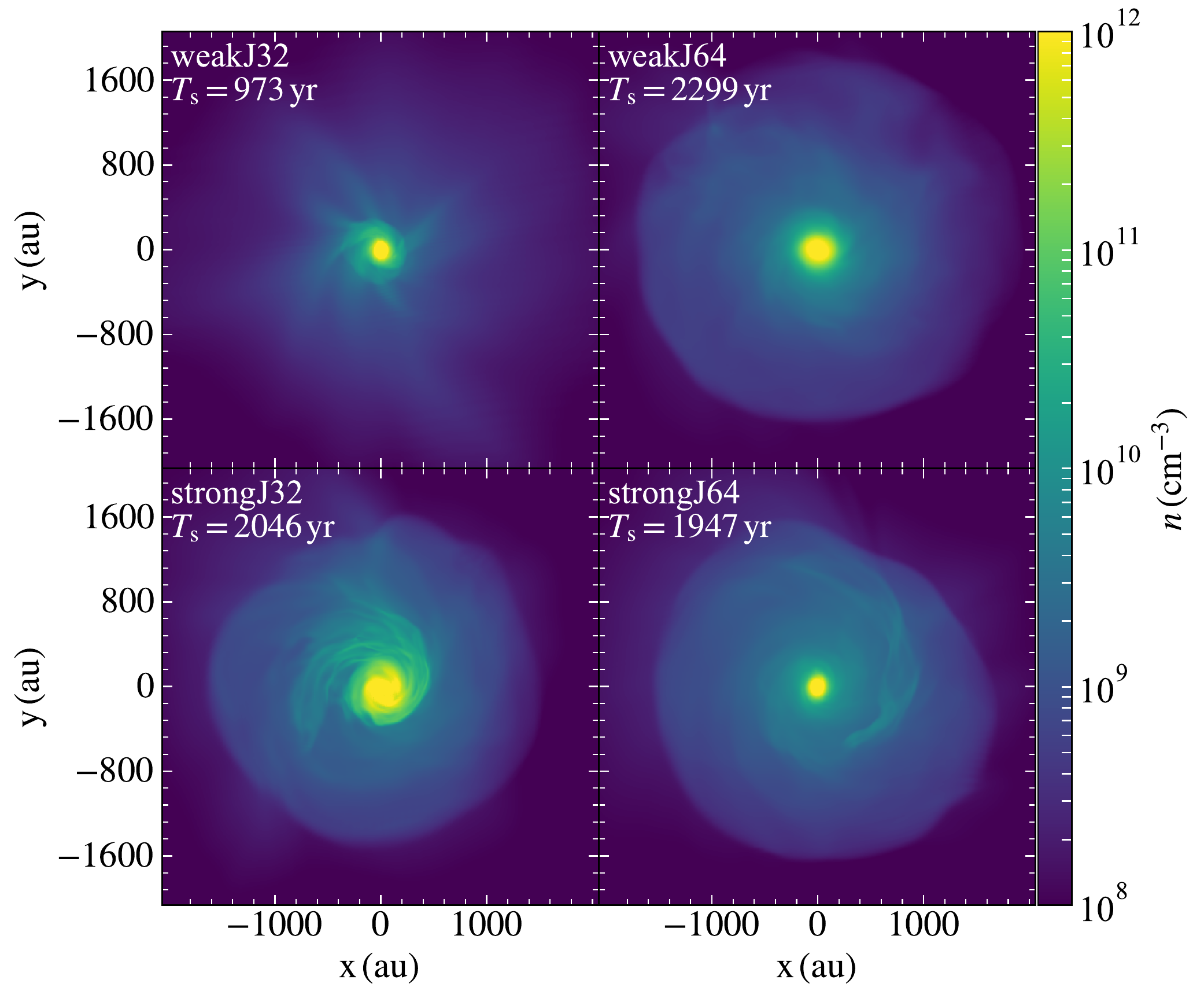}
\includegraphics[width=\columnwidth]{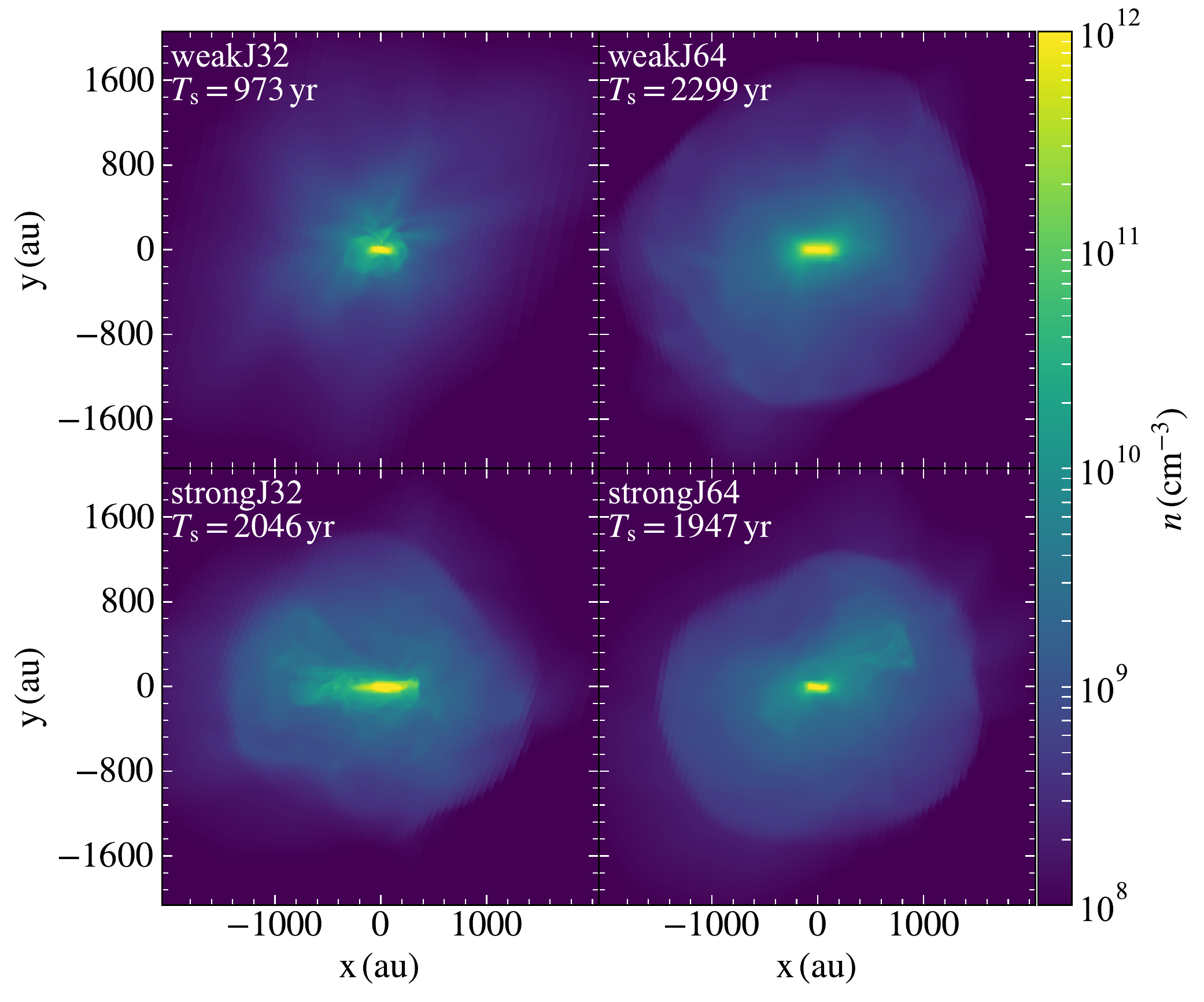}
\includegraphics[width=\columnwidth]{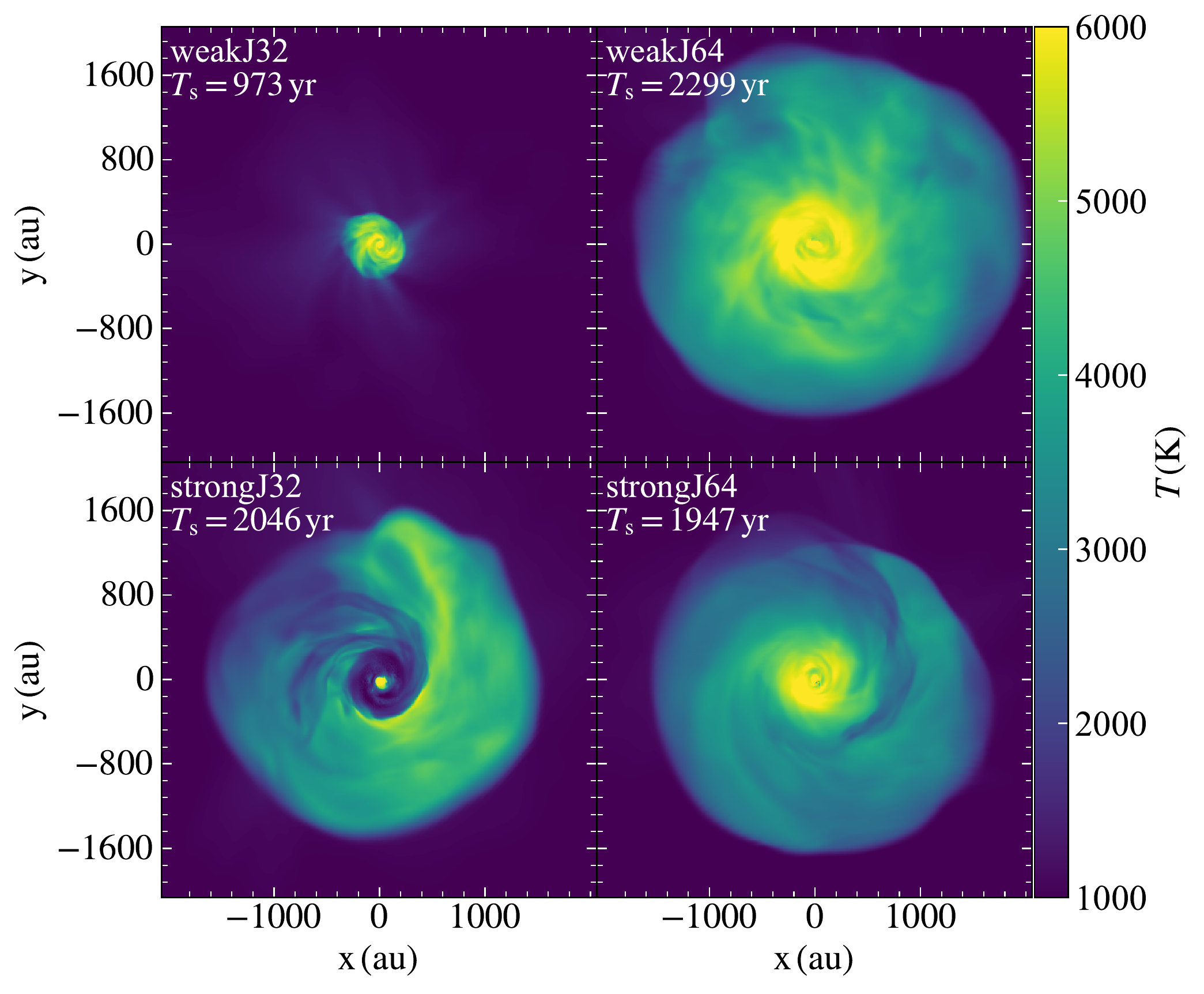}
\includegraphics[width=\columnwidth]{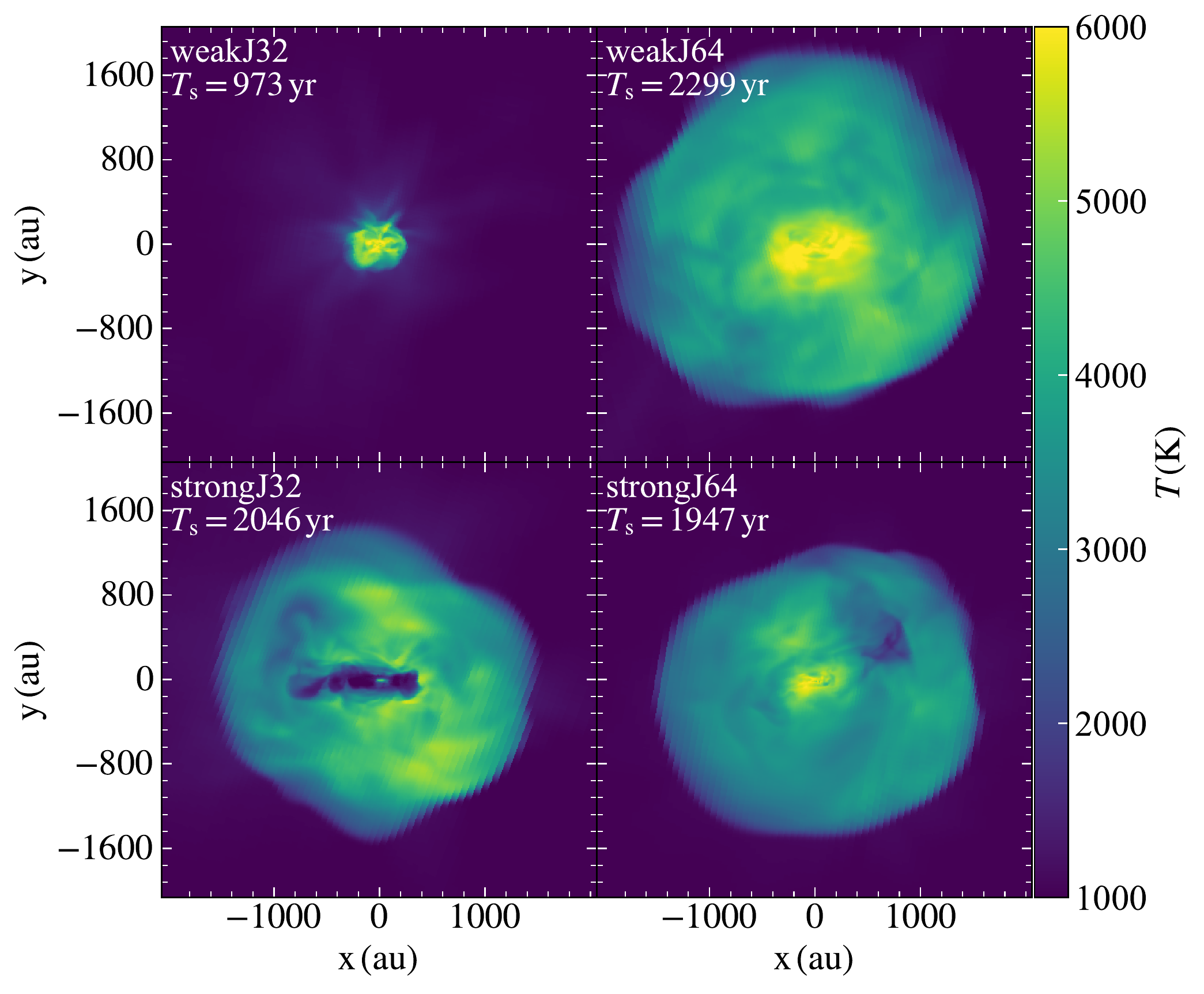}
\includegraphics[width=\columnwidth]{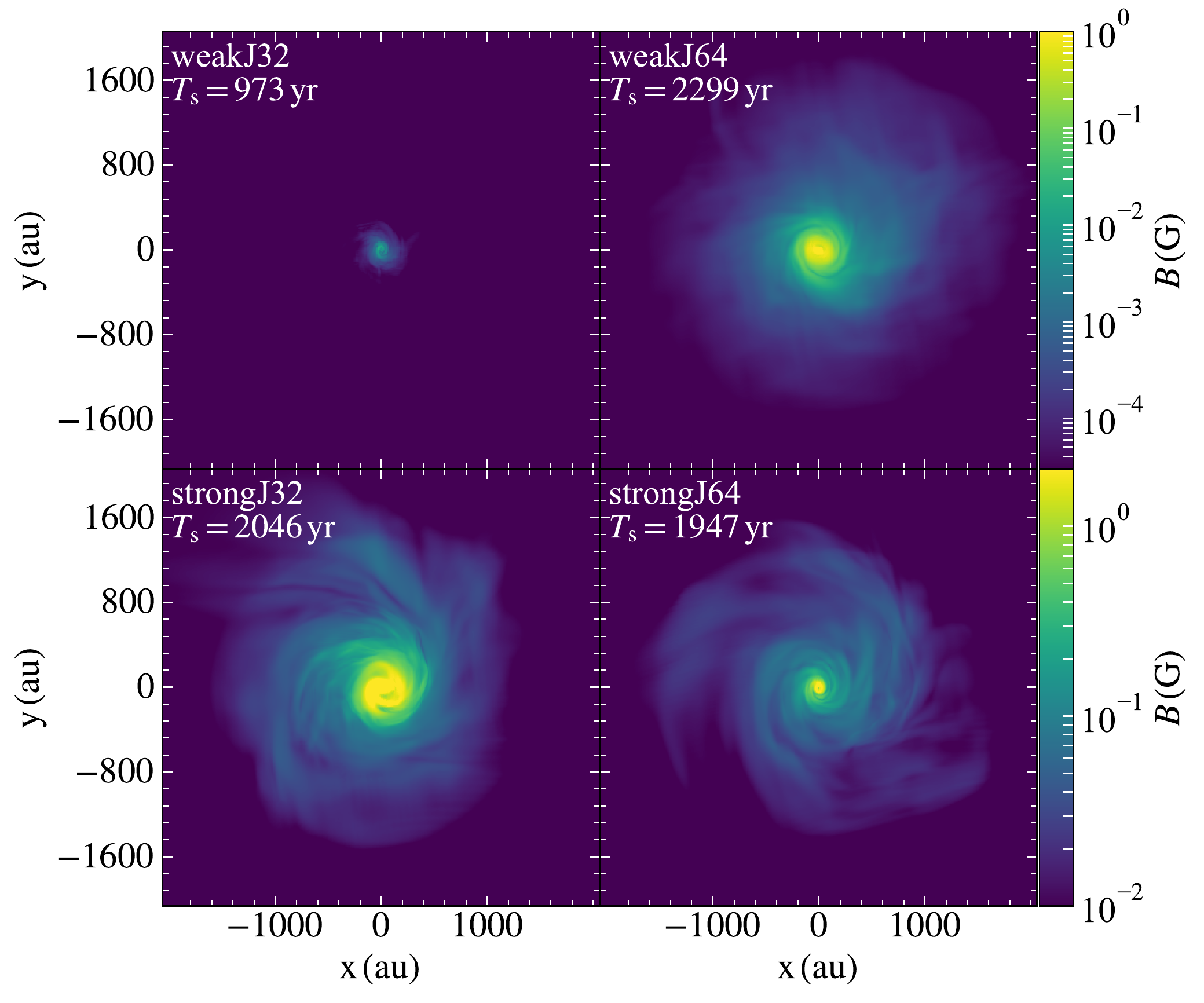}
\includegraphics[width=\columnwidth]{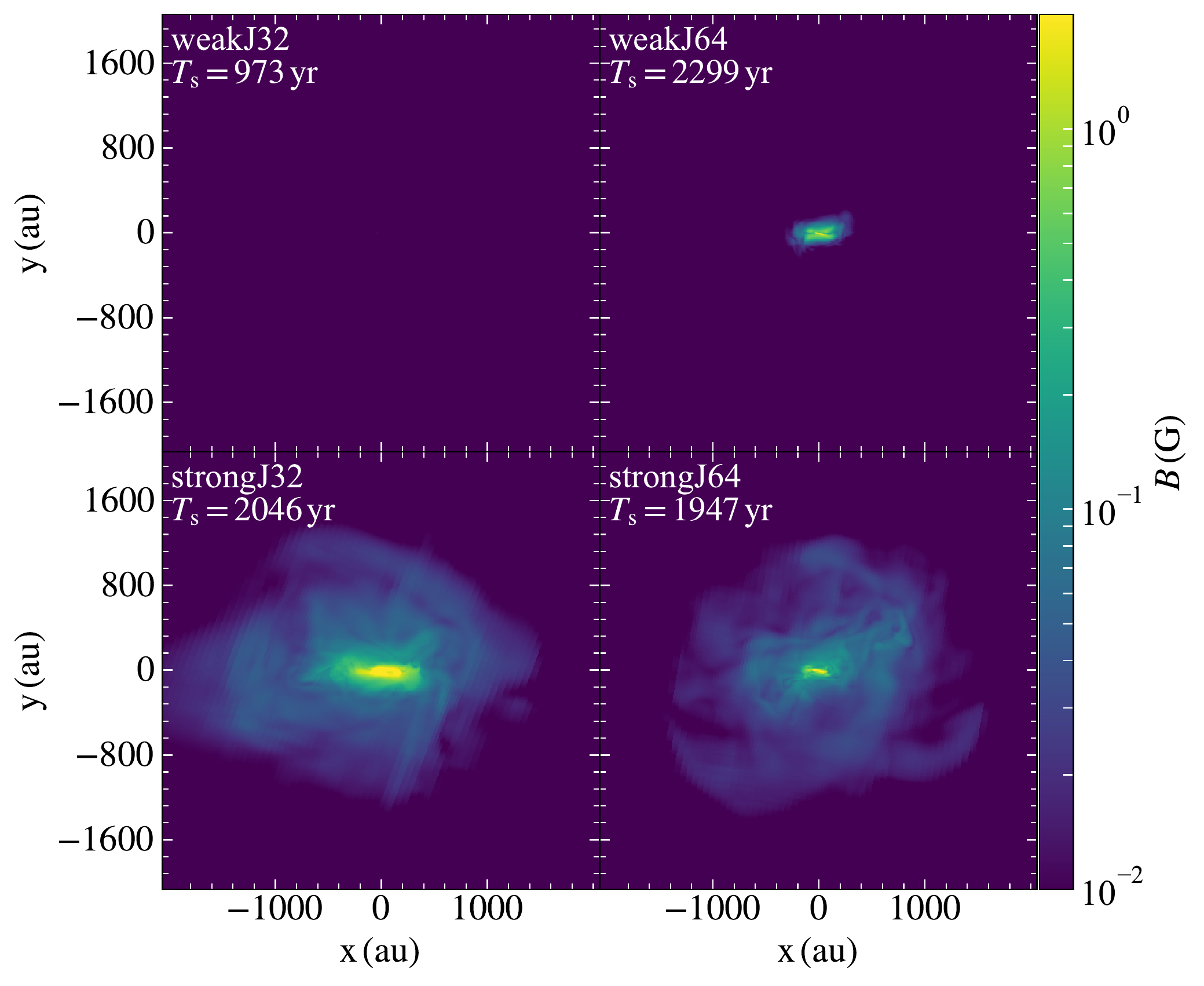}
\caption{Density-weighted face-on (left) and edge-on (right) projections of the number density $n$ (top panels), temperature $T$ (middle panels), and magnetic field strength $B$ (bottom panels) centred on position of the sink particle, for the four different categories of runs we study in this work (see \autoref{tab:tab1}). The snapshots correspond to the end of the simulation where SFE = 5 percent and the sink particle has accreted $50\,\mathrm{M}_{\odot}$. The simulations shown differ only in resolution and initial magnetic field strength.}
\label{fig:proj_dens}
\end{figure*}

\begin{figure*}
\includegraphics[width=\linewidth]{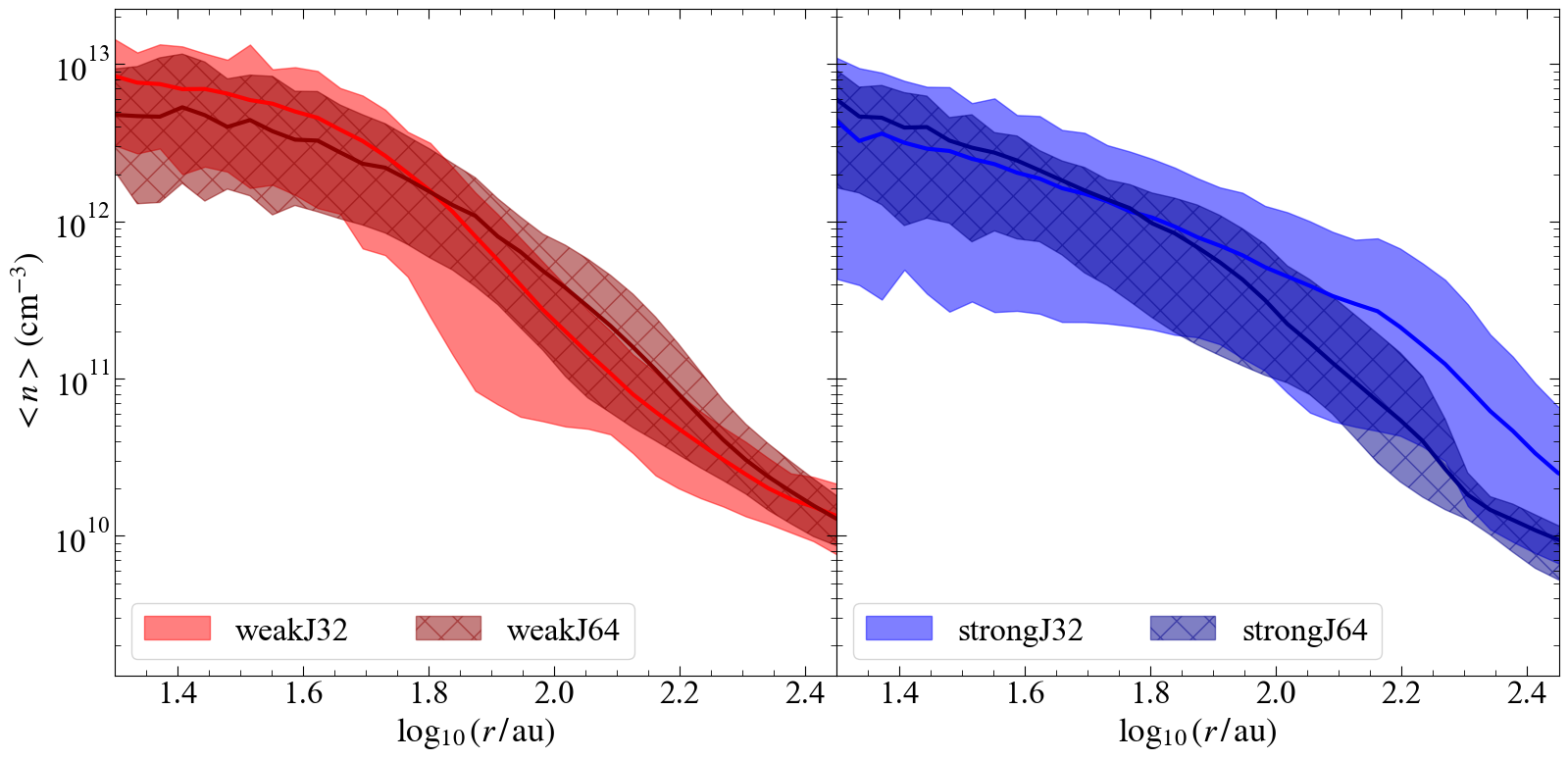}
\caption{Azimuthally-averaged, mass-weighted radial profiles of the number density in the accretion discs around the central star for all the non-fragmenting set of realisations (\textit{i.e., } where only a single sink particle forms), shown at the end of the simulation when SFE = 5 percent (see \autoref{s:analysis_discussions}). The four sets of simulations denoted in the legend represent weak and strong magnetic fields run with 32 and 64 cells per Jeans length (see \autoref{tab:tab1}). The solid curves represent the mean averaged over all the non-fragmenting realisations in each simulation category. The coloured bands mark the $5^{\rm{th}}$ to $95^{\rm{th}}$ percentile range. Note that the radial extent of these profiles only covers the accretion discs, and is smaller than the extent of the projections we show in \autoref{fig:proj_dens}.}
\label{fig:thermprofiles}
\end{figure*}

\begin{figure*}
\includegraphics[width=\linewidth]{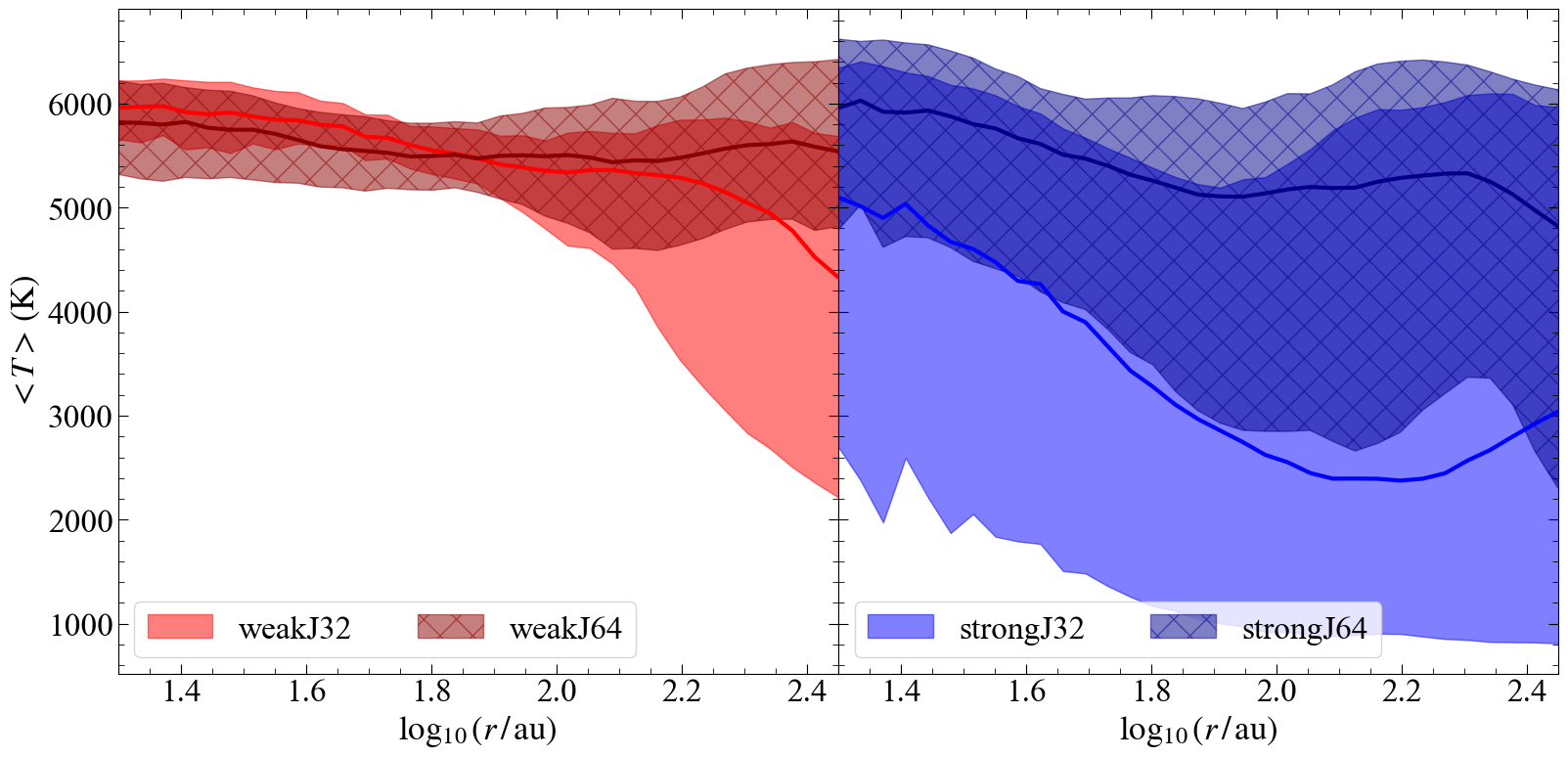}
\caption{Same as \autoref{fig:thermprofiles} but for the temperature in the discs.}
\label{fig:thermprofiles2}
\end{figure*}

\subsection{Morphology and thermal evolution}
\label{s:projections}
We begin our discussion by examining the morphology and thermal structure of the discs formed in four representative realisations, one from each of the combinations of resolution and magnetic field strength listed in \autoref{tab:tab1}. \autoref{fig:proj_dens} shows the face-on and edge-on projections of density-weighted number density ($n$), temperature ($T$) and magnetic field strength ($B$) for these four example realisations. All the snapshots are centred at the single sink particle that forms in the simulations (noting again that for this part of the analysis we select simulations that only form one star), and show the time at which the simulation reaches SFE = 5 percent. $T_{\mathrm{s}}$ denotes the time elapsed since the formation of the sink particle. It is straightforward to notice that the morphology of the system varies significantly in the run weakJ32 as compared to the other three runs. In runs weakJ64, strongJ32 and strongJ64, the snapshots reveal the presence of a hot, spherical bubble that expands radially outwards with time (see movies M1 and M2 attached as online material with this paper for reference) such that there are higher temperatures inside the bubble that lead to more dissociation of \hii. A similar resolution-dependent effect has been noted by \citet{2012ApJ...745..154T} during the formation and collapse of dark matter minihaloes in their cosmological simulations with a seed magnetic field. However, this phenomenon does not occur primarily due to magnetic fields. We show in \aref{s:app_thermalbubble} that the qualitative difference in the outcome is a result of how well we resolve the length and timescales for chemical evolution and radiative cooling across shock fronts. However, the effect is not related to dynamo amplification, and has little impact on the overall results because the thermal pressure is dynamically-unimportant in all cases. For this reason, we do not discuss it further in the main text.

In addition to the hot spherical halo that extends across $\sim 2000\,\rm{au}$, the face-on and edge-on projections in \autoref{fig:proj_dens} also reveal the presence of an accretion disc a few $100\,\rm{au}$ in size (typically $\lesssim 500\,\rm{au}$ in diameter) in each run. The fact that discs form in our ideal MHD simulations even with strong magnetic fields implies the absence of the magnetic braking catastrophe, and could be a result of the misalignment between the rotation axis and the magnetic field, as has been found in Population I star formation \citep[e.g.,][]{2013MNRAS.432.3320S,2013A&A...554A..17J,2013ApJ...774...82L,2018MNRAS.473.2124G}. Movies M1 and M2 in the supplementary material show the 3D orientation and geometry of the disc around the star in each of the four runs. The mass of accretion discs around the $50\,\rm{M_{\odot}}$ sink particle varies between $5-40\,\rm{M_{\odot}}$ within the different non-fragmenting runs, with a mean disc mass $\sim 16\,\rm{M_{\odot}}$ in all the four categories listed in \autoref{tab:tab1}. We also find that discs in the J64 runs are systematically more massive than the discs in the J32 runs irrespective of the magnetic field strength.

The difference that is of the greatest interest to us is in the magnetic fields (see bottom panel of \autoref{fig:proj_dens}). Strikingly, we see that the magnetic field strength and morphology of the weakJ64 run is much closer to the results we find for strongJ32 or strongJ64 than to weakJ32. Despite having started from identical initial conditions, the field in weakJ64 is $\sim 3$ orders of magnitude stronger than in weakJ32. We explore this in detail in the following subsections.

Unless explicitly stated otherwise, we calculate all quantities of interest in the frame of reference of the disc once it is formed, averaging over a cylindrical region centred on the sink particle, with the symmetry axis of the cylinder aligned with the angular momentum vector of the mass within 500 au of the sink particle. We define the usual cylindrical basis vectors $(\hat{r},\hat{\phi},\hat{z})$ to denote position within this analysis region. We find that using an analysis region of diameter $500\,\mathrm{au}$ and half-height $50\,\mathrm{au}$ ensures that the resulting volume covers the entire disc in all our realisations. We have also verified that our results are relatively insensitive to the exact choice of radius and height for our analysis region (provided it is large enough to cover most of the mass of the disc), since we calculate mass-weighted quantities, which means that the low-density material does not contribute significantly to our quantitative analyses of the disc material.

\subsection{Radial profiles}
\label{s:radialprofiles}
We next discuss the statistical properties of the discs formed in every non-fragmenting realisation. In \autoref{fig:thermprofiles} and \autoref{fig:thermprofiles2} we show the azimuthally-averaged, mass-weighted radial profiles of the number density and the temperature in accretion discs at the end of the simulation where the SFE reaches 5 per cent. The solid curves show the mean value averaged over all the realisations we include, and the coloured bands denote the $5^{\rm{th}}$ to $95^{\rm{th}}$ percentile range. The profiles of number density in the accretion discs are quite similar in all the four runs, with somewhat larger scatter in the strong magnetic field runs. On the contrary, the radial profiles of temperature show significant differences, between strong and weak fields, as well as between runs with 32 and 64 cells per Jeans length. Simulations with strong magnetic fields show a much larger scatter in the disc temperatures as compared to simulations with the weak field.

\begin{figure}
\includegraphics[width=0.95\columnwidth]{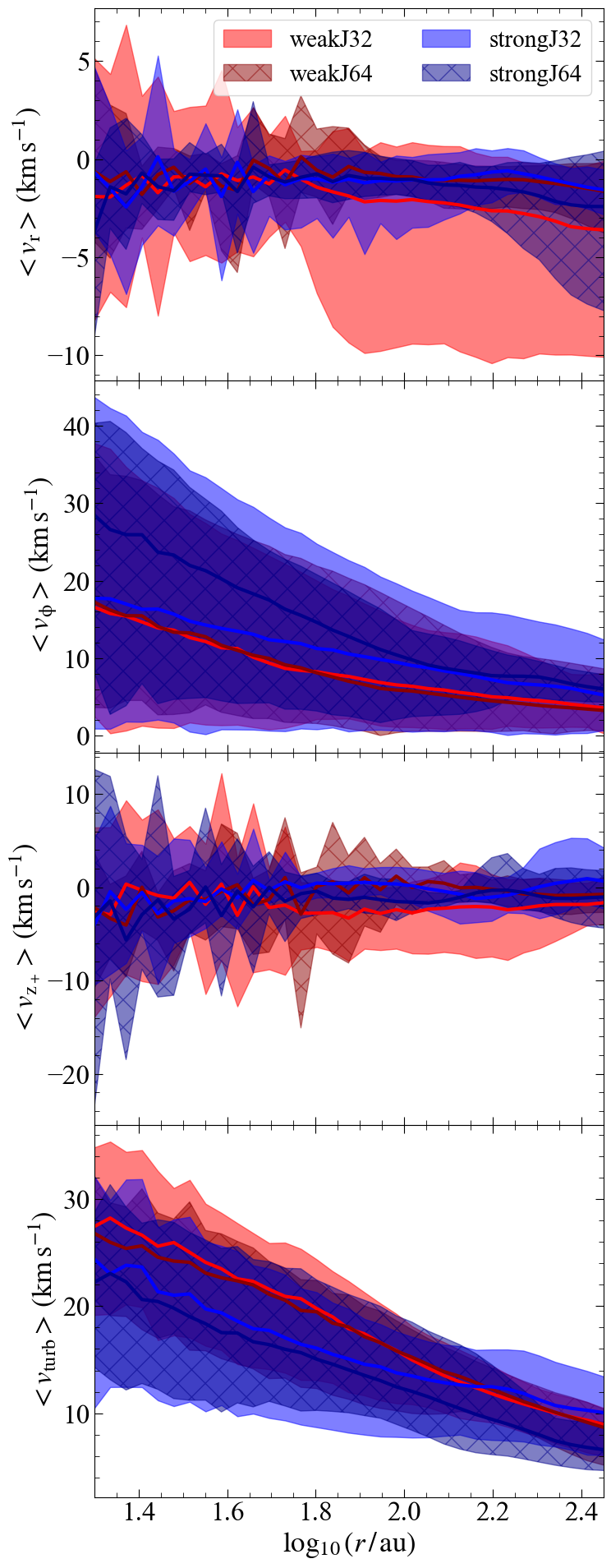}
\caption{Azimuthally-averaged, mass-weighted radial profiles of different components of velocity in the accretion discs around the central star for all the non-fragmenting realisations in each set of simulations, shown at the end of the simulation when SFE = 5 percent. $v_{\mathrm{z_+}}$ refers to the velocity component along the polar axis of the disc in its upper half (see \autoref{s:radialprofiles} for details). $\langle v_{\mathrm{turb}} \rangle$ in the last panel is the turbulent component of the velocity, defined in \autoref{eq:vturb}. Solid curves represent the mean value and coloured bands represent the $5^{\rm{th}}$ to $95^{\rm{th}}$ percentile range.}
\label{fig:vprofiles}
\end{figure}

In \autoref{fig:vprofiles}, we plot the azimuthally-averaged, mass-weighted radial profiles of the different components of the velocity field in the accretion discs. We define the turbulent component of the velocity, $\mathbf{v}_{\mathrm{turb}}$, as,
\begin{equation}
\mathbf{v}_{\mathrm{turb}} = (v_{\mathrm{r}} - \langle v_{\mathrm{r}} \rangle)\,\hat{r} + (v_{\mathrm{\phi}} - \langle v_{\mathrm{\phi}} \rangle)\,\hat{\phi} + (v_{\mathrm{z_{+/-}}} - \langle v_{\mathrm{z_{+/-}}} \rangle)\,\hat{z}\,,
\label{eq:vturb}
\end{equation}
where $v_{\rm{r}}$ and $v_\phi$ are the radial and toroidal cylindrical components of velocity, and $v_{\mathrm{z_+}}$ ($v_{\mathrm{z_-}}$) denotes the component of velocity along the $\hat{z}$ axis of the disc in its upper (lower) half. Angle brackets indicate the azimuthal average of a given quantity; we denote the magnitude of the turbulent field as $v_{\rm turb}\equiv \left|\mathbf{v}_{\mathrm{turb}}\right|$. We subtract the mean components of $v_{\mathrm{z_+}}$ and $v_{\mathrm{z_-}}$ separately because $\langle v_{\mathrm{z}} \rangle \approx 0$; this systematically removes the influence of any radial or vertical gradients and gives a reasonable estimate of the turbulent velocity \citep{2011ApJ...735..122F}. We see that in all cases our discs are dominated by azimuthal and turbulent motions, with much smaller radial inflow and vertical infall velocities. There is no obvious systematic difference in the results with either resolution or initial magnetic field strength. 

\begin{figure}
\includegraphics[width=0.95\columnwidth]{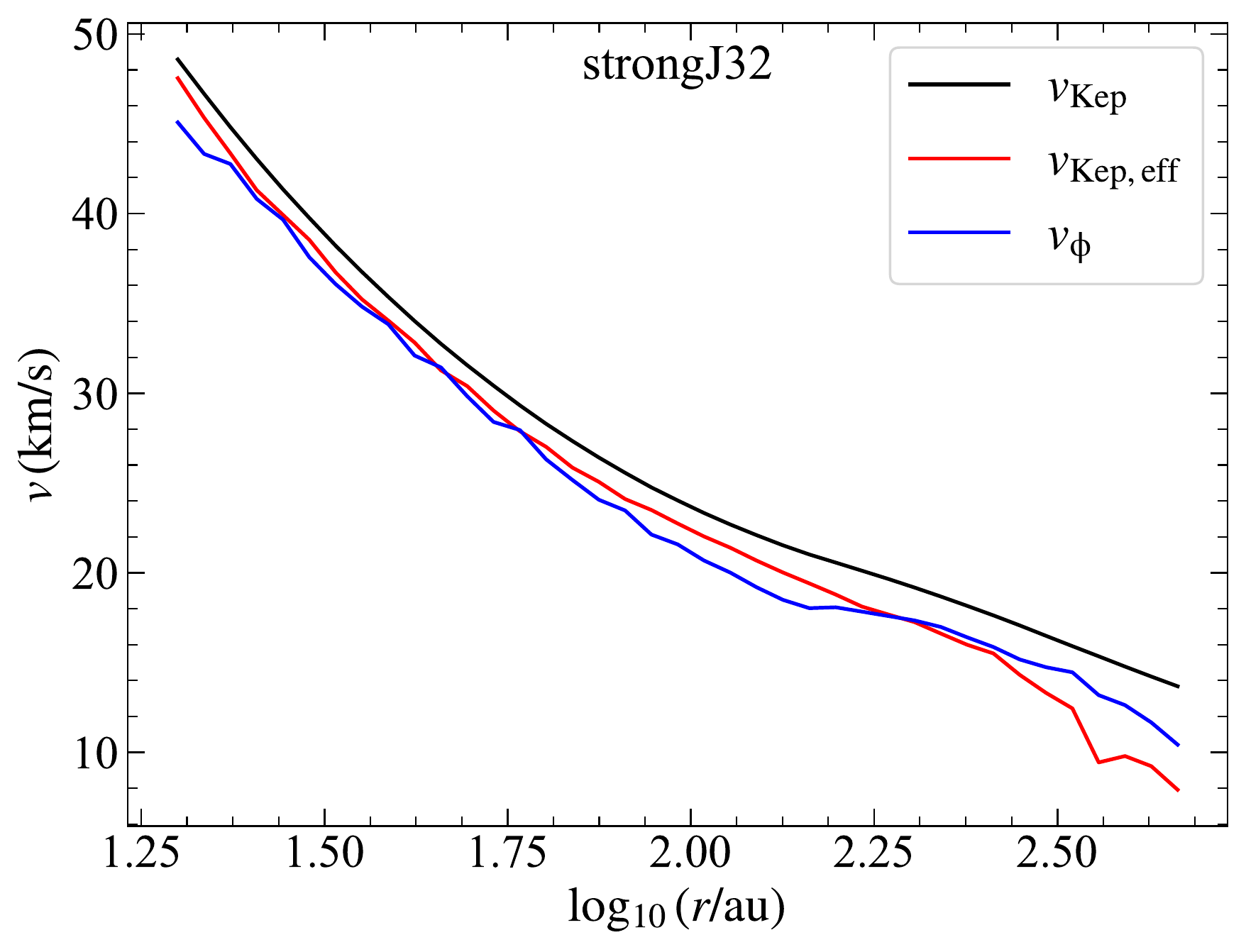}
\caption{Comparison of the toroidal component of the velocity ($v_{\phi}$, blue) with the Keplerian velocity ($v_\mathrm{Kep}$, black) in the disc for a representative realisation of the simulation strongJ32. The effective Keplerian velocity ($v_{\rm{Kep,\,eff}}$, red) is obtained by subtracting the contribution due to the Lorentz force from the Keplerian velocity as defined in \autoref{eq:veff_kep}. The disc is slightly sub-Keplerian due to additional support from magnetic pressure.}
\label{fig:keplereff}
\end{figure}

Because differential rotation is important for large-scale dynamo action, it is important to characterise the degree of shear in our discs, which is given by $q = -\mathrm{dln}\,(v_{\mathrm{\phi}}/r)/\mathrm{dln}\,r$. For a disc in rigid (non-differential) rotation, $q = 0$, whereas for a Keplerian disc, $q = 1.5$. Using the profiles of $v_{\mathrm{\phi}}$ from \autoref{fig:vprofiles}, we find $q \approx 1.4$. Thus, the discs are almost Keplerian, but there is some deviation from a perfect Keplerian disc, which is due to additional support from magnetic pressure. To demonstrate this, we define an effective Keplerian velocity $v_{\mathrm{eff,Kep}}$ in the presence of magnetic fields by subtracting the contribution of the Lorentz force,
\begin{equation}
\frac{v^2_{\mathrm{eff,Kep}}}{r} =  \frac{v^2_{\mathrm{Kep}}}{r}- \frac{1}{4\pi\rho}\,[(\nabla \times B) \times B]\,,
\label{eq:veff_kep}
\end{equation}
where $v_{\mathrm{Kep}} = \sqrt{GM_{\mathrm{s}}/r}$, $M_{\rm{s}}$ is the mass of the sink particle, and $\rho$ is the gas density; the final term represents the Lorentz acceleration. As an example, we plot the radial profile of $v_{\mathrm{eff,Kep}}$ for one of the strongJ32 runs in \autoref{fig:keplereff}. We see that the $v_{\mathrm{\phi}}$ is almost exactly $v_{\mathrm{eff,Kep}}$ at all radii, and and the deviation from $v_{\rm Kep}$ is $\sim 10\%$. Thus we find that our discs are in near-Keplerian rotation, with a small deviation from a Keplerian profile due to magnetic support.

\begin{figure}
\includegraphics[width=0.95\columnwidth]{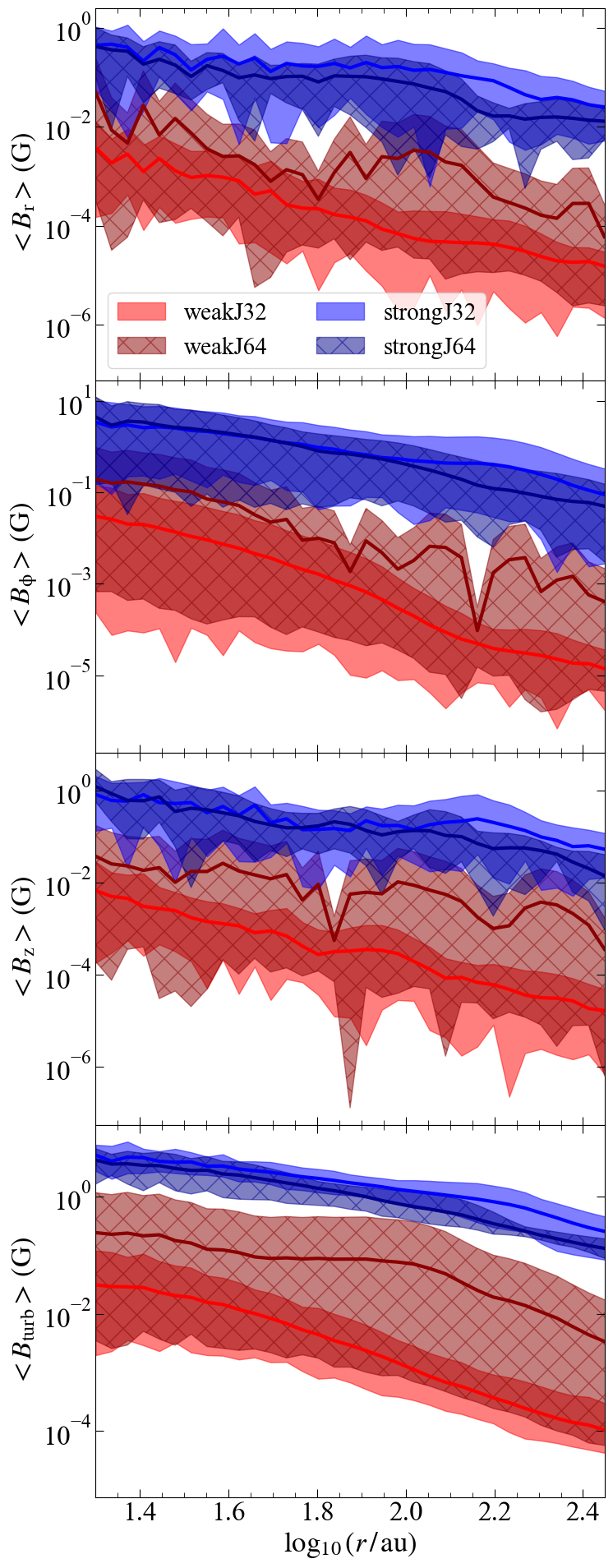}
\caption{Same as \autoref{fig:vprofiles}, but for the different components of the magnetic field. $B_{\mathrm{turb}}$ is defined as in \autoref{eq:Bturb}. Note that $\langle B_{\mathrm{\phi}} \rangle$ is the largest component, indicating a large-scale mean field in the toroidal direction. There is also a strong turbulent component, $\langle B_{\mathrm{turb}} \rangle$, indicating the presence of the small-scale dynamo.}
\label{fig:Bprofiles}
\end{figure}

In \autoref{fig:Bprofiles} we plot mass-weighted, azimuthally-averaged radial profiles of the different components of the magnetic fields. In analogy to $\mathbf{v}_{\mathrm{turb}}$, we define the turbulent component of the field, $\mathbf{B}_{\mathrm{turb}}$, as,
\begin{equation}
\mathbf{B}_{\mathrm{turb}} = (B_{\mathrm{r}} - \langle B_{\mathrm{r}} \rangle)\,\hat{r} + (B_{\mathrm{\phi}} - \langle B_{\mathrm{\phi}} \rangle)\,\hat{\phi} + (B_{\mathrm{z}} - \langle B_{\mathrm{z}} \rangle)\,\hat{z}\,,
\label{eq:Bturb}
\end{equation}
where $B_{\rm{r}}$, $B_\phi$, and $B_{\rm{z}}$ are the cylindrical components of the total magnetic field, and we denote the magnitude of the turbulent field as $B_{\rm turb}\equiv \left|\mathbf{B}_{\mathrm{turb}}\right|$. In line with the morphological differences between weakJ32 and the other runs, we find that all the components of the field are substantially lower in weakJ32 compared to the other runs. We also see that while the initial magnetic field we imposed is completely random, in all cases except weakJ32, a substantial mean toroidal field develops in the disc, as is clear from the radial profile of $\langle B_{\mathrm{\phi}} \rangle$ in \autoref{fig:Bprofiles}. This component is comparable in strength to the turbulent component.

By looking at the time evolution of the velocity and the magnetic field profiles (available as movie M3 in the supplementary material), we find that initially, when the sink forms, all the three components of the velocity and the magnetic field are of the same strength. As the disc around the sink starts to grow and expand outwards to conserve angular momentum, a strong toroidal component of velocity ($v_{\phi}$) is generated, which winds up the magnetic field in the $\hat{\phi}$ direction, thus giving rise to a strong $B_{\phi}$ component. This happens through the development of the $\Omega$ effect that results from shear instabilities \citep{1961ApJ...133..572B}. We explore the $\Omega$ effect further in \autoref{s:magfield_amp_largescale}.

\subsection{Magnetic field amplification}
\label{s:magfield_amp}
We have seen that in our weakJ64 simulations starting from an initially weak field, the simulations eventually develop both strong turbulent and mean fields. This suggests the operation of both the small-scale turbulent and the large-scale mean-field dynamo in the disc. Here, we attempt to separate the dynamos based on the scale of turbulent motions that amplify the small-scale dynamo and the scale at which inhomogeneties exist in a mean field generated through the large-scale dynamo \citep{2018JPlPh..84e7302R}. In principle, scale separation cannot be naively applied here, because we only have $\sim70$ resolution elements across the disc, and we do not resolve the innermost part of the disc where we have less than 30 resolution elements across the disc scale height. Nevertheless, we have an effective magnetic Reynolds number in the discs of the order of $300-600$ (based on $(2r/\Delta x)^{4/3-3/2}$, where $\Delta x=7.6\,\rm{au}$, and the exponents correspond to Kolmogorov and Burgers scaling, respectively; see \citealt{2011ApJ...731...62F}), which is sufficient to trigger the small-scale dynamo for magnetic Prandtl numbers close to unity \citep{2004PhRvE..70a6308H}, but we cannot capture its full efficiency in our simulations. This is not unexpected, given that the viscous and Ohmic dissipation scales that control the driving scale and the growth rate of the dynamo lie $2-3$ orders of magnitude below the current resolution that can be afforded by simulations like ours \citep[Figure~12]{2019MNRAS.488.1846N}. Similarly, since we resolve the scale height in the inner disc with less than 30 resolution elements, it means that we cannot fully capture the large-scale dynamo amplification. Resolving the dynamo action not only depends on the absolute resolution, but also on the Jeans resolution, such that increasing the Jeans resolution leads to more efficient dynamo amplification \citep{2010ApJ...721L.134S,2011ApJ...731...62F}. Thus, scale separation is possibly also a function of the Jeans resolution. Keeping these caveats in mind, we caution that we may be in a transitionary regime in between the small-scale and the large-scale dynamo since we do not fully resolve the action of either dynamo. In fact, as \cite{2005PhR...417....1B} note, the differentiation between a small-scale and a large-scale dynamo is artificial, and in reality, the two regimes are connected. It is believed that the small-scale dynamo acts first, possibly even before the disc is established, but the disc is needed for the large-scale dynamo to act. 

In the next two subsections, we quantify the action of these dynamos in accretion discs around the sink particles in our simulations. We remind the reader that we only use those simulations that did not show any fragmentation ($\sim$ 8 out of the total 25 simulations in each category) and formed only a single sink particle of mass $50\,\rm{M_{\odot}}$ by SFE = 5 per cent, while studying the small-scale and the large-scale dynamo. Analysing multiple runs provides the benefit of overcoming statistical noise between different runs \citep[e.g.,][]{2019MNRAS.490..513S,2020arXiv200211502S,2020MNRAS.494.1871W}, but our choice of which simulations we include implies that we only study dynamo amplification in accretion discs around massive first stars.

\subsubsection{Small-Scale dynamo}
\label{s:magfield_amp_dynamo}
Traditionally, the presence of a small-scale dynamo is verified by an exponential increase in the ratio,
\begin{equation}
Q_{\mathrm{ss}} = \frac{(B_{\mathrm{turb}})_{\mathrm{rms}}}{\rho^{2/3}}\,,
\label{eq:smallscaledynamo}
\end{equation}
over the lifetime of the simulation (e.g., \citealt{2010ApJ...721L.134S,2011ApJ...731...62F,2012ApJ...745..154T,2012ApJ...754...99S,2013MNRAS.432..668L,2015PhRvE..92b3010S,2016JPlPh..82f5301F}); here, $\left(B_{\rm turb}\right)_{\rm rms}$ is the root-mean-square strength of the turbulent component of the magnetic field, averaged over some region of interest (see below). The motivation for the normalisation by $\rho^{2/3}$ in the definition of $Q_{\rm ss}$ is to remove the effects of flux-freezing: even in the absence of dynamo action, a collapse that increases the gas density will also increase the strength of the frozen-in field. The fastest growth occurs for the spherical collapse of a region with a dynamically-unimportant, tangled field, in which case $B\propto \rho^{2/3}$ \citep{2006ApJ...641..949B,2010ApJ...725..466C}; stronger fields that force anisotropic collapse produce scalings closer to $B\propto \rho^{1/2}$ \citep{1973ApJ...182..387A,1999ApJ...520..706C,2001ApJ...550..314D,2004ApJ...605..800L,2006ApJ...647L...1M,2017ApJ...838...40M,2019FrASS...6....5H}. Thus, $Q_{\rm ss}$ is either a conserved or decreasing quantity in the absence of dynamo action, and an increase in $Q_{\rm ss}$ indicates that the small-scale dynamo is operating.

\begin{figure*}
\includegraphics[width=\linewidth]{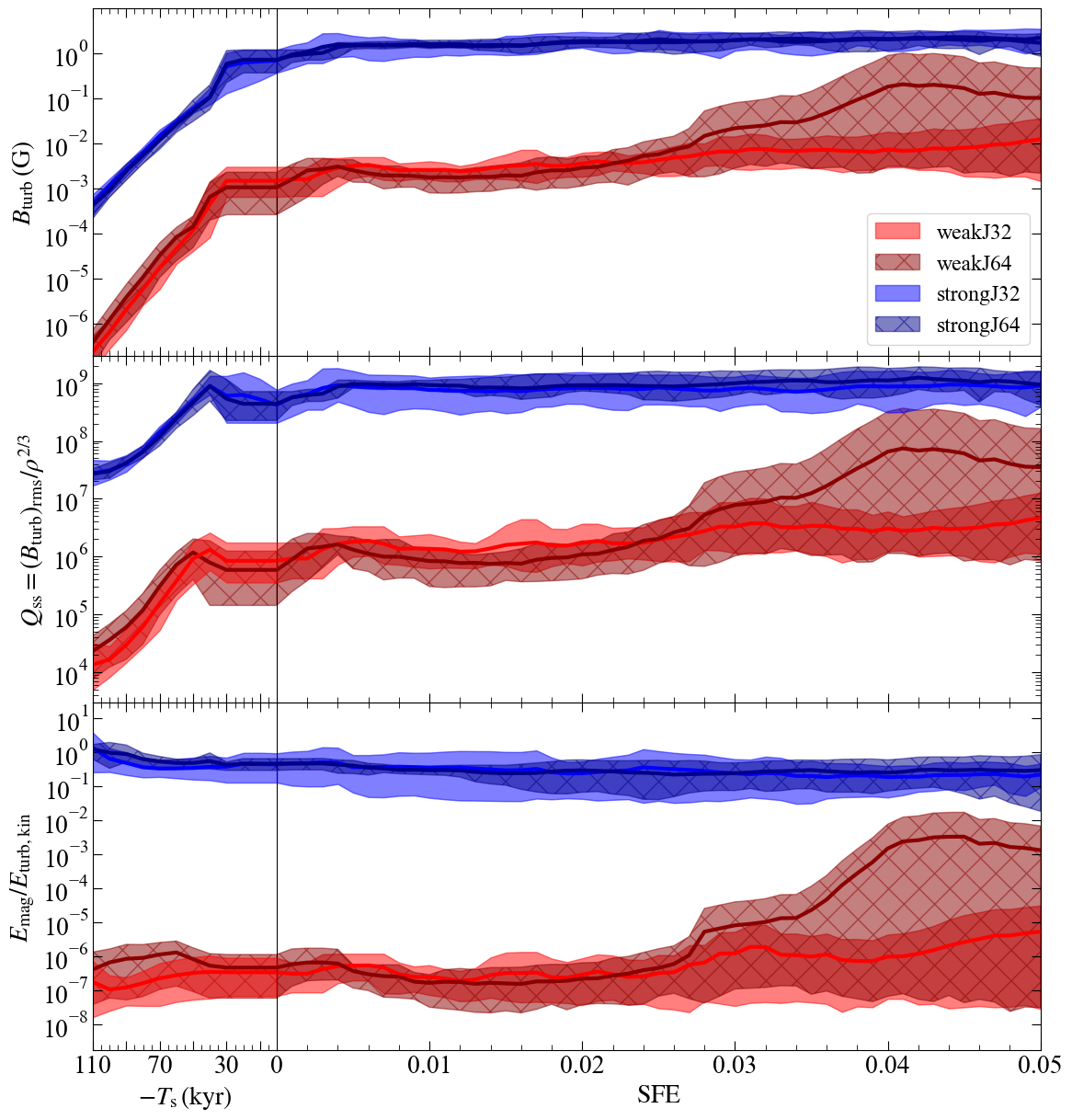}
\caption{\textit{Top panel:} Evolution of the mass-averaged turbulent magnetic field $B_{\rm{turb}}$ as a function of time in the core before the formation of the sink at time $T_{\mathrm{s}}$, and as a function of star formation efficiency (SFE) in the disc around the sink after its formation (SFE = 0.05 implies that the sink particle has accreted $50\,\mathrm{M}_{\odot}$). We calculate $B_{\mathrm{turb}}$ using \autoref{eq:Bturb}, averaging over a spherical volume of radius $0.01\,\mathrm{pc}$ before the collapse, and a cylindrical region of radius $500\,\mathrm{au}$ and half-height $50\,\mathrm{au}$, oriented to lie in the same plane as the accretion disc, afterwards. The solid lines represent the mean averaged over the non-fragmenting ($N_{\rm{r}}\sim 8$) realisations of each set of simulations with weak and strong magnetic fields at two different Jeans resolution as marked in the legend (see also, \autoref{tab:tab1}). The coloured bands represent the $5^{\mathrm{th}}$ to $95^{\mathrm{th}}$ percentile range. \textit{Middle panel:} The evolution of the small-scale dynamo ratio, $Q_{\mathrm{ss}}$, calculated using \autoref{eq:smallscaledynamo}. \textit{Bottom panel:} The bottom panel shows the ratio of magnetic to turbulent kinetic energy, which quantifies the growth and saturation of the small-scale dynamo.}
\label{fig:dynamo}
\end{figure*}

We show the value of $B_{\rm{turb}}$ and $Q_{\rm ss}$ versus time for all of our non-fragmenting runs in the top and middle panels of \autoref{fig:dynamo}. For the purpose of this plot, we calculate all quantities in a spherical region of radius $0.01\,\mathrm{pc}$ centred on the point of maximum density before the sink particle forms, and then shift to a cylindrical geometry that represents the accretion disc around the sink, which we defined in \autoref{s:projections}. However, our results are quite insensitive to these choices, as long as the volume over which we compute $Q_{\rm ss}$ is large enough to capture the entire disc. In \autoref{fig:dynamo}, the solid lines are the mean values averaged over the $\sim$ 8 non-fragmenting simulations in each category, and the colored bands denote the $5^{\mathrm{th}}$ and the $95^{\mathrm{th}}$ percentiles\footnote{The percentiles requested can be outside the range that can be computed given the limited input sample size in our work. To take this into account, we use the \texttt{numpy} percentile function with the \texttt{linear} interpolation option such that if the request percentile is between two data points $i$ and $j$, this operation returns $i + (j - i)\times f$ , where $f$ is the fractional part of the sample index between $i$ and $j$; see the \texttt{numpy} user manual for further details \citep{oliphant2006guide}.}.

The initial amplification in $Q_{\rm{ss}}$ in the pre-sink phase ($T_{\mathrm{s}} < 0$) is not accompanied by an increase in the ratio of the magnetic to the turbulent kinetic energy, $E_{\mathrm{mag}}/E_{\mathrm{turb,kin}}$, as shown in the bottom panel of \autoref{fig:dynamo}. Here, we define the turbulent kinetic energy as $E_{\rm turb} = \sum (1/2)m v_{\rm turb}^2$, where the sum is over all cells in the region of interest, $m$ is the cell mass, and $v_{\rm turb}$ is the turbulent velocity as defined in \autoref{eq:vturb}. $E_{\mathrm{mag}}/E_{\mathrm{turb,kin}}$ remains constant because the ongoing collapse converts gravitational potential energy into turbulent motions and thence into magnetic fields \citep[e.g.,][]{2020ApJ...899..115X}, resulting in an increase in both $E_{\mathrm{mag}}$ and $E_{\mathrm{turb,kin}}$. This is a collapse-driven dynamo, as observed earlier by \cite{2010ApJ...721L.134S} and \cite{2011PhRvL.107k4504F}. There is a small plateau close to the sink formation time, $T_{\mathrm{s}} = 0$, which results because the evolution is so fast that the snapshots we use (which are taken every 50 timesteps) do not resolve the time frames that we parameterize by the SFE.

Turning now to the phase of the simulation after sink formation, the plot shows that, on average, the weakJ64 runs show a substantial small-scale dynamo amplification. The value of $Q_{\rm ss}$ asymptotically approaches the value found in the strong-field runs. However, there is a large scatter, so the amount of dynamo amplification varies significantly with the random seed for the initial turbulent velocity and magnetic field. On the other hand, runs with an initially strong magnetic field do not show any amplification in $Q_{\mathrm{ss}}$, independent of resolution. This is in accordance with the expectations laid out in section~2 of \citetalias{2020arXiv200211502S}, namely that the strong-field runs correspond to an initially saturated magnetic field that cannot be further amplified. 

Consistent with our discussion of $Q_{\rm ss}$, we see that the ratio $E_{\mathrm{mag}}/E_{\mathrm{turb,kin}}$ is nearly constant in the strong-field runs, further implying that the field is saturated. The saturation level is around $0.001-0.1$, in very good agreement with that expected from isothermal MHD turbulence simulations with similar Mach number and magnetic Prandtl number \citep{2011PhRvL.107k4504F,2014ApJ...797L..19F,2016JPlPh..82f5301F}, but here with realistic chemistry and cooling. Most interestingly, in the weakJ64 case, the ratio of energies increases from $\sim 10^{-7}$ for our initial state to $\sim 10^{-3} - 10^{-2}$ by the time the SFE has reached $4-5$ per cent. However, there is a great deal of scatter about this result, with some runs showing no increase in magnetic energy density at all, and others reaching a ratio of almost $0.1$.

\begin{figure}
\includegraphics[width=\columnwidth]{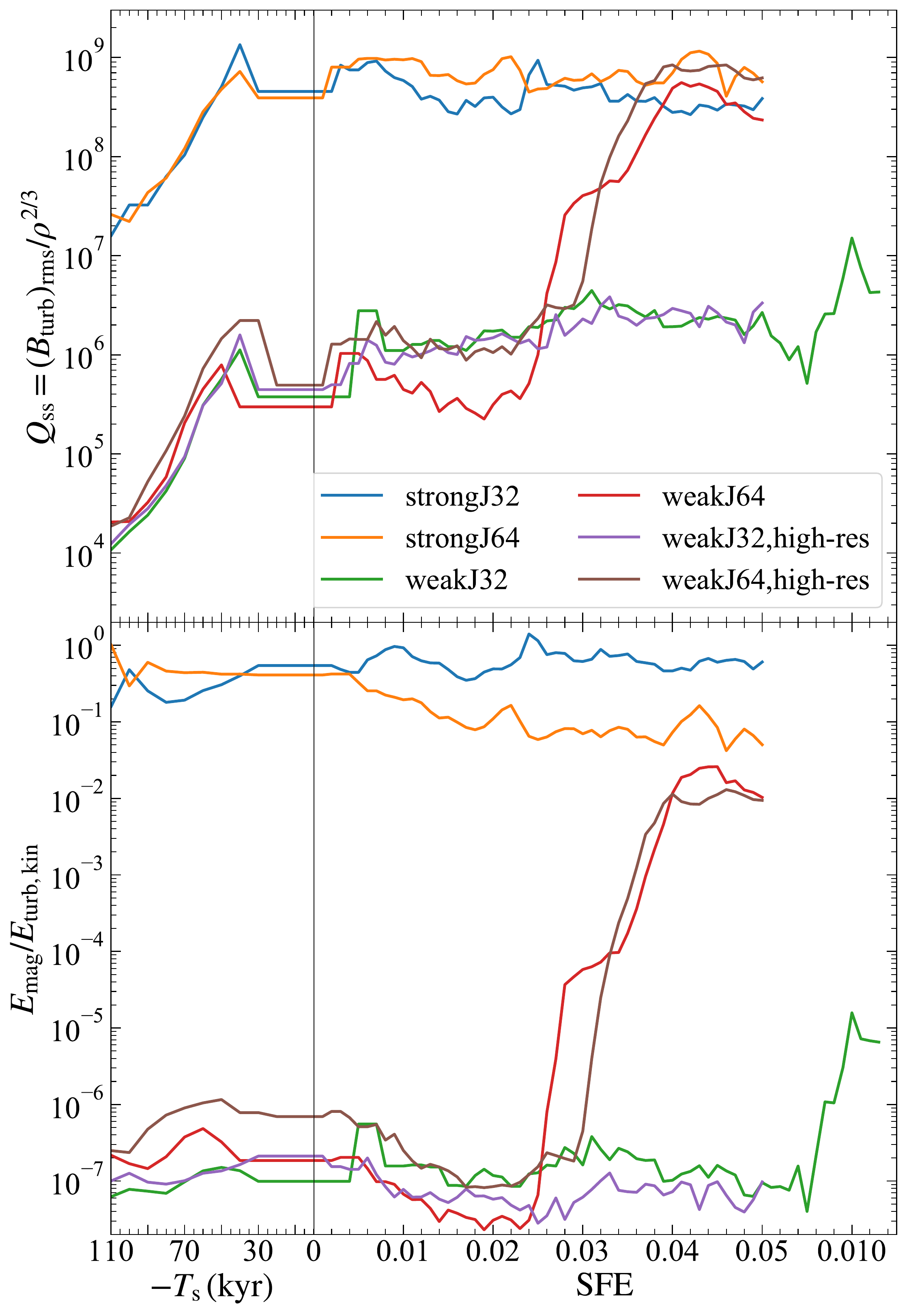}
\caption{Same as \autoref{fig:dynamo}, but for one particular realization, including runs with the weak field at a higher absolute resolution. In this plot, the strongJ32, strongJ64, and weakJ64 cases are all run with the standard resolution. We run the weakJ32 case shown with the standard resolution as well, but allow the run to continue to SFE = 12 percent rather than 5 percent. Finally, for the two runs (weakJ32, high-res) and (weakJ64, low-res), we use the same initial conditions and refinement criteria as weakJ32 and weakJ64, but add an extra level of refinement, so the maximum resolution is $\Delta x = 3.8$~au rather than 7.6~au. The main conclusion from this is that higher Jeans resolution is more critical for resolving dynamo amplification than absolute maximum resolution.}
\label{fig:dynamo_one}
\end{figure}

While it may seem from \autoref{fig:dynamo} that the small-scale dynamo action is not resolved with 32 cells per Jeans length, this is not strictly the case. In fact, field amplification is only delayed, not suppressed entirely. To illustrate this point, we have continued one realisation of a weakJ32 run to an SFE of 12~percent; we show $Q_{\mathrm{ss}}$ and $E_{\rm mag}/E_{\rm turb,kin}$ for this run in \autoref{fig:dynamo_one}\footnote{We caution that the evolution at this point is largely unphysical, because we are not including stellar radiation feedback, which would be extremely important for a 120 $M_\odot$ star as it forms in this case; we should therefore think of this run as a numerical experiment to demonstrate a point about dynamo action, rather than a realistic simulation of the formation of a primordial star.}. As the green curve in the top panel of \autoref{fig:dynamo_one} shows, small-scale dynamo amplification does occur, but not until after SFE = 5 percent. Thus, the small-scale dynamo is active even at a J=32 Jeans resolution; however the time at which amplification begins seems to be both stochastic and resolution-dependent. This observation confirms that $\mathrm{J}\sim30$ is a threshold for dynamo amplification \citep{2010ApJ...721L.134S,2011ApJ...731...62F} even in the presence of primordial chemistry and cooling.

We also use this realisation to test for the effects of increasing the maximum resolution, as opposed to changing the number of cells per Jeans length. To this end, we repeat the weak-field case with 32 and 64 cells per Jeans length but at a higher absolute resolution, such that $\Delta x = 3.8\,\mathrm{au}$ on the finest AMR level (instead of the $\Delta x = 7.6\,\mathrm{au}$ for all the other simulations). It is clear from \autoref{fig:dynamo_one} that the runs with higher absolute resolution produce results that are very similar to the ones at our standard absolute resolution. While we are unable to repeat these higher-resolution tests in more cases due to the computational expense, the experiment we have performed suggests that absolute resolution is less important for capturing small-scale dynamo effects than resolving the Jeans length by a sufficiently large number of cells. Further, we also find that the onset of the small-scale dynamo action depends on  the degree of smoothness and circularity in the disc. We show this in the movie M4, by comparing the evolution of magnetic field strength in two realizations of the weakJ64 runs that show no and high amplification, respectively. This demands a detailed analysis of the interaction of disc dynamos with disc instabilities, which is beyond the scope of this work since the inner disc is not well resolved, as we discuss in \autoref{s:magfield_amp_largescale}.

\subsubsection{Large-Scale dynamo}
\label{s:magfield_amp_largescale}
The kinetic helicity, $F = \int v\cdot W dV$ (where $W = \nabla \times v$ is the vorticity) is finite and non-zero in our simulations, thus suggesting the presence of helical turbulence (e.g., \citealt{1999ARA&A..37...37K,2005PhR...417....1B,2019PhRvF...4b4608B}). It is well known that helical turbulence in the presence of a vertical density gradient (stratification) and differential rotation in discs can lead to the generation of a large-scale magnetic field through the $\alpha\Omega$ dynamo \citep{1981MNRAS.195..881P,1981MNRAS.195..897P}. While the small-scale dynamo generates field structures on smaller scales, it cannot lead to the production of a coherent field on large scales. The presence of the mean toroidal field as we observe in our simulation hints at the presence of a large-scale dynamo. This happens due to winding-up of the magnetic field in the toroidal direction by shearing motions ($\Omega$ effect, see \citealt{1961ApJ...133..572B}). However, the $\Omega$ effect alone cannot be classified as a dynamo. The amplified poloidal component that we observe in addition to the toroidal field implies that an additional field amplification is mechanism at work, likely the $\alpha$ effect \citep{1966ZNatA..21..369S}. The combination of the $\alpha$ and $\Omega$ effects is well known to give rise to the $\alpha\Omega$ large-scale dynamo \citep{2005PhR...417....1B}. In our simulations, we speculate that the $\alpha\Omega$ dynamo acts to amplify the small-scale field produced by the small-scale dynamo (provided the resolution is high enough), and that this transforms the small-scale field into the large-scale one that we observe. While it is generally believed that the small-scale dynamo can quench the action of the mean-field dynamo \citep{1992ApJ...396..606K,1999PhRvL..83.2957S,2004ApJ...612..276S,2005PhR...417....1B,2012SSRv..169..123B}, recent high-resolution simulations find that a large-scale mean field can co-exist with a small-scale field of comparable strength, if both shear and helical turbulence are present \citep{2016MNRAS.461..240B,2017ApJ...850L...8S,2019arXiv190508278B}, due to the unified action of the two dynamos. Note that the studies we refer to above have quite different simulation setups that the one we use; for example, most of these studies assume incompressibility and isotropy, work with low Mach numbers, do not have density stratification, and do not include other relevant astrophysical processes like gravity, non-equilibrium chemistry, and non-isothermal thermodynamics; thus, they do not form or study the dynamo in a star-forming environment. Hence, results from these studies cannot be directly compared against our simulations, and the extent to which they apply to our work is limited. Nevertheless, these studies provide important basic insight and a reasonable starting point to discuss dynamo operation in astrophysical discs around the first stars.

The operation of the $\alpha$ effect depends on the competition between how efficiently the field is regenerated as compared to how quickly is it dissipated (by turbulence) in the poloidal direction. Similarly, the operation of the $\Omega$ effect depends on how efficiently the field is amplified as compared to how quickly is it dissipated in the toroidal direction. Thus, the two effects can be quantified under the assumption of axisymmetric accretion discs \citep{1986ESASP.251..569R} by taking the ratio of field amplification rate to its dissipation rate \citep{1981MNRAS.195..897P,1988ASSL..133.....R,1990ApJ...362..318S,2005PhR...417....1B},
\begin{equation}
R_{\alpha} = \frac{\alpha h}{\eta_{\mathrm{T}}}\,\mathrm{and}\,\,R_{\Omega} = \frac{Sh^2}{\eta_{\mathrm{T}}}\,
\label{eq:R_alphaRomega}
\end{equation}
where $h$ is the disc scale height at some radius $r$ and $S$ is the radial shear caused by differential rotation, $S = r\,\partial\Omega/\partial r$. Further, $\alpha$ is a pseudo-scalar\footnote{The pseudo-scalar, $\alpha$, is actually a compressed version of the symmetric part of the $\alpha$ tensor, obtained under the assumption that the turbulent field is isotropic (invariant under rotation) and homogeneous \citep[see equation~7.15 in][]{1978magnetic}. Certain simulations have calculated the different components of the $\alpha$ tensor (e.g., \citealt{2007GApFD.101...81S,2018A&A...609A..51W,2019ApJ...886...21V,2020MNRAS.491.3870B}), however, as we explain in the main text, this is not within the scope of this work.} that represents the transport coefficient responsible for the $\alpha$ effect ($\alpha = 0$ if the turbulence is not helical), and $\eta_{\mathrm{T}}$ is the second transport coefficient, given as the sum of microscopic and turbulent magnetic diffusivity \citep{1978magnetic,1980opp..bookR....K,1988Natur.336..341R,2018JPlPh..84d7304B}. Theoretically, the operation of the large-scale dynamo requires that the large-scale dynamo number,
\begin{equation}
D_{\alpha\Omega} = R_{\alpha} R_{\Omega}\,,
\label{eq:dynamo_number}
\end{equation}
be larger than unity, implying that the amplification of the field by the two effects is more rapid than dissipation\footnote{In practice, the critical dynamo number above which the dynamo operation is sustained is a function of the disc aspect ratio \citep[see their Figure 2]{2019Galax...7...91B}, however, it is generally expected to be of the order of $1-10$ in astrophysical systems with disc-like geometries \citep{1988ASSL..133.....R,1988Natur.336..341R}.}. 

In order to verify that a large-scale $\alpha\Omega$ dynamo is operating in our simulations, we must estimate $\alpha$ and $\eta_{\rm{T}}$, so that we may compute $R_\alpha$ and $R_\Omega$, and hence $D_{\alpha\Omega}$ (\autoref{eq:dynamo_number}). For accretion discs, the microscopic diffusivity is much less than the turbulent magnetic diffusivity as the discs are highly conducting (e.g., \citealt{1976JMP....17.1808K,1981MNRAS.195..881P,1998ApJ...495..385H}; see, however, \autoref{s:nonidealMHD}, where we discuss the effects of non-ideal MHD, \textit{i.e.,} the effects of microscopic diffusivities primarily giving rise to ambipolar diffusion). In a simulation such as ours, which does not include explicit resistivity and where the physical scale of magnetic diffusion is unresolved, the magnetic diffusivity is dictated solely by the finite resolution of the grid on which we discretise the MHD equations \citep{2009ApJ...700...63K,2012ApJ...747...21S,2020arXiv200614607M}. We can estimate the diffusivity by noting that, in the absence of explicit viscosity or resistivity, the dissipation scale is always of order the cell size $\Delta x$, and thus the fluid and magnetic Reynolds numbers Re and Rm must be close to unity for length scales $\ell\sim \Delta x$ (e.g., \citealt{2004PhRvE..70a6308H,2004ApJ...612..276S,2004ApJ...617..339B}), giving a magnetic Prandtl number around 1 \citep{2007MNRAS.381..319L,2011PhRvL.107k4504F,2020arXiv200614607M}. Thus, $\eta_{\mathrm{T}} \sim c_{\mathrm{s}} \Delta x \sim 10^{20}\,\mathrm{cm^2\,s^{-1}}$ (see, however, \autoref{s:nonidealMHD}).

To calculate $\alpha$, we make use of the fact that, in the presence of helical turbulence, the induction equation for the mean field has an additional term, $\chi$, that depends on the turbulent velocity and magnetic field \citep[see their equation~151]{2016RPPh...79g6901S}. Assuming spatially isotropic turbulence and a finite scale separation between small and large scales \citep{2002PhRvL..89z5007B}, $\chi$ can be expressed under a first-order smoothing approximation (neglecting quadratic terms) in the kinematic regime as,
\begin{equation}
\chi = \langle \mathbf{v}_{\mathrm{turb}} \times \mathbf{B}_{\mathrm{turb}} \rangle  = \alpha \langle \mathbf{B} \rangle - \eta_{\mathrm{T}}\,\nabla \times \langle \mathbf{B} \rangle\,.
\label{eq:induction_meanfielddynamo}
\end{equation}
Note that \autoref{eq:induction_meanfielddynamo} can only be used if: (1) Rm is small (\citealt{2009MNRAS.395L..48C}; see, however, \cite{2013Natur.497..463T} and \cite{2014ApJ...789...70C} who show that the large-scale dynamo can persist even when Rm is high, provided there exists a strong shear), and (2) $B_{\mathrm{turb}}$ is small compared to $\langle B \rangle$. The latter assumption is violated in our simulations, since $B_{\mathrm{turb}} \sim \langle B \rangle$. However, direct numerical simulations report that \autoref{eq:induction_meanfielddynamo} holds approximately even when $B_{\mathrm{turb}} \sim \langle B \rangle$ \citep{2008MNRAS.385L..15S}, especially in the case of accretion discs, because the turbulence correlation time is small compared to the turnover time \citep{1981MNRAS.195..881P,2005PhR...417....1B,2019JPlPh..85d2001R}. Since our goal is not to estimate an accurate value of $\alpha$, but simply to check if the $\alpha$ effect operates in our simulations, we work under the first-order smoothing approximation introduced above. Plugging $\eta_{\rm{T}}$ into \autoref{eq:induction_meanfielddynamo} gives $\langle \alpha \rangle \approx 3\,\mathrm{km\,s^{-1}}$. Note that we derive $\eta_{\mathrm{T}}$ (and by extension, $\alpha$) based on the grid resolution. Nonetheless, the values we obtain are in very good agreement with that expected from the first-order smoothing approximation for $\langle v_{\mathrm{turb}} \rangle \sim 10\,\mathrm{km\,s^{-1}}$ as in our simulations \citep[see their equation~16]{2008MNRAS.385L..15S}\footnote{If Rm $\leq$ 1, the first-order smoothing approximation estimates have to be scaled by Rm.}.

\begin{figure*}
\includegraphics[width=\linewidth]{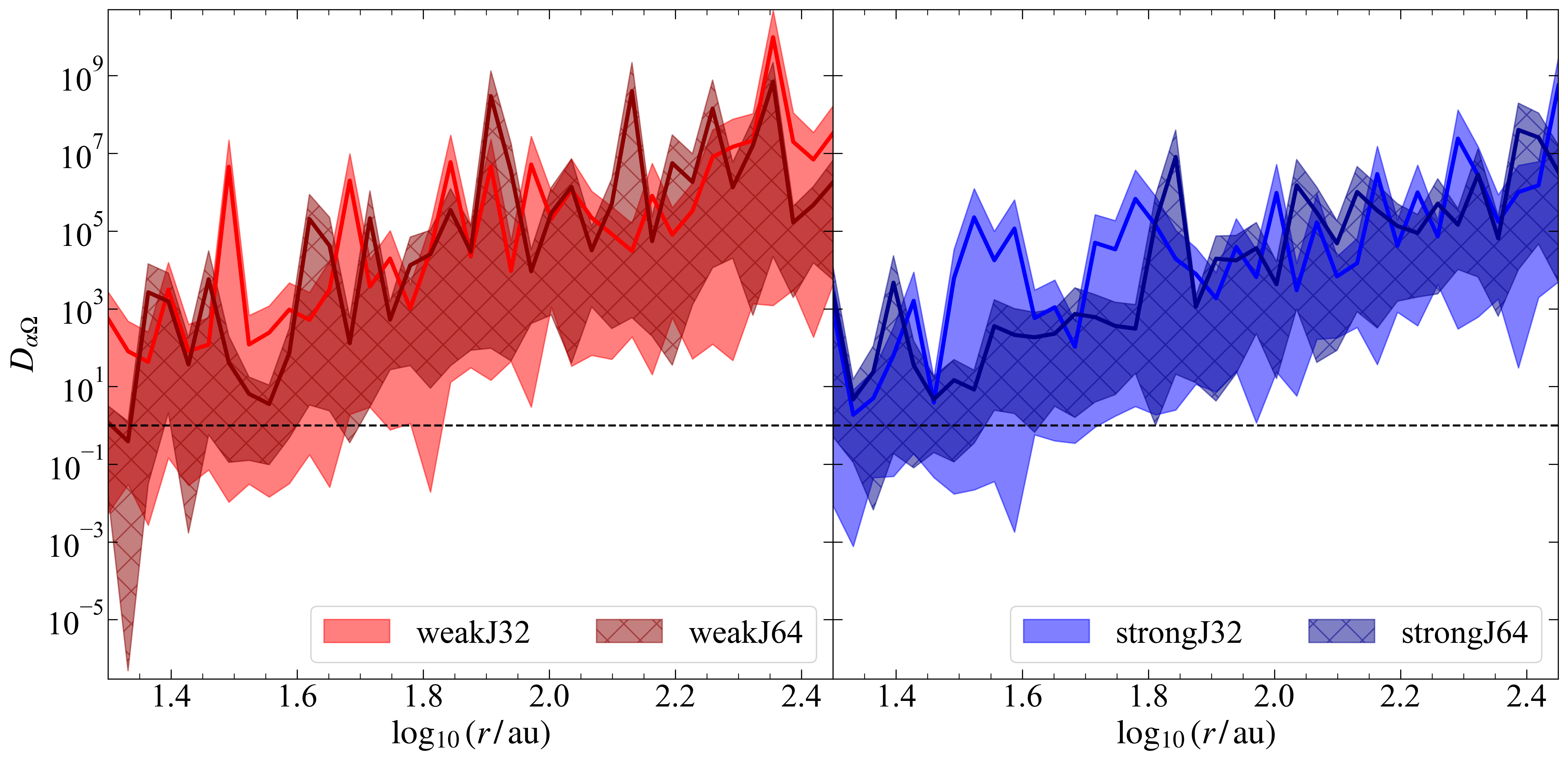}
\caption{Azimuthally-averaged radial profile of the large-scale dynamo number, $D_{\alpha\Omega}$ (see \autoref{eq:dynamo_number}), in the disc for different runs at SFE = 5 percent. The left and right panels and present results for weak and strong magnetic field cases with different Jeans resolution, respectively (cf. \autoref{tab:tab1} for details). Similar to \autoref{fig:dynamo}, the solid lines represent the mean over all simulations that produced a single sink particle, and the colored bands represent the $5^{\rm{th}}$ and $95^{\rm{th}}$ percentiles. This mean-field dynamo likely operates due to the $\alpha\Omega$ effect in the disc, requiring a critical $D_{\alpha\Omega} > 1$, denoted by the dashed, black line. It does not act in the inner disc in some realisations due to coarser resolution there, but for $\log_{10}(r/\mathrm{au})\gtrsim1.5$, the weak-field models have $D_{\alpha\Omega}>1$ and all models have $D_{\alpha\Omega}\gg1$ further out in the disc ($r\gtrsim100\,\mathrm{au}$), demonstrating the effectiveness of the $\alpha\Omega$ dynamo. However, note that we are likely overestimating the value of $D_{\alpha\Omega}$ in the case of strong fields since we do not include non-ideal MHD effects that can dissipate the field (see \autoref{s:nonidealMHD}).}
\label{fig:dynamo_number}
\end{figure*}

Our results confirm the recent results of \citet{2019arXiv191107898L}, who also argue for the presence of a large-scale mean-field dynamo acting in Population III star formation. However, we note that \citeauthor{2019arXiv191107898L} used only 8~cells per Jeans length in their simulations, which is not sufficient to capture the small-scale dynamo \citep{2011ApJ...731...62F}. Thus, they likely miss the production of small-scale fields that can then be driven to large scales by the $\alpha\Omega$ effect. In addition, both \citeauthor{2019arXiv191107898L}'s simulations and ours likely underestimate the rate of $\alpha\Omega$ dynamo amplification because, as we show above, the dynamo number $D_{\alpha\Omega} \propto (h/\eta_{\rm T})^2$, which for a simulation dominated by grid dissipation ($\eta_{\rm T}\propto \Delta x$), implies a scaling $D_{\alpha\Omega}\propto (h/\Delta x)^2$. In practice, this means that in order to capture the $\alpha\Omega$ effect well requires that the disc scale height be resolved by at least $\sim 30$ cells \citep{2011ApJ...731...62F}. We approach, but do not quite satisfy this requirement in the outer disc, and fall far short of it at smaller radii where the disc is thinner. We remind the reader that due to coarse resolution in the inner disc, we can thus only qualitatively comment on the scale separation between the small and the large-scale dynamo. We also point out that our analysis implies that the growth rate of the $\alpha\Omega$ dynamo depends on the absolute resolution, not just the number of cells per Jeans length.

Finally, we note that there can be additional large-scale dynamo amplification in the presence of helical turbulence and strong shear, for e.g., the shear current effect \citep{2003PhRvE..68c6301R,2004PhRvE..70d6310R} or the incoherent $\alpha$-shear dynamo \citep{1988ApJ...332..857H,1997ApJ...475..263V}. We have not explored these effects in this work, so we cannot rule out the possibility that they might be operating as well.

\subsection{Effects of non-ideal MHD}
\label{s:nonidealMHD}
The results we have discussed so far are based on ideal MHD simulations, and it is important to highlight the effects of non-ideal processes on the operation of the dynamo. \cite{2019MNRAS.488.1846N} use one-zone calculations to study the ionisation fraction and the effects of non-ideal MHD on primordial chemistry, and conclude that non-ideal processes do not suppress dynamo amplification when the field is weak, but they can have a significant impact in the case of strong fields. This is because at high densities ($\sim\, 10^{10}-10^{13}\,\rm{cm^{-3}}$) in the presence of strong fields ($\sim\! 1\,\rm{G}$), magnetic diffusivities giving rise to ambipolar diffusion can be as high as $10^{21}-10^{23}\,\rm{cm^2\,s^{-1}}$, which is $\sim 1-3$ orders of magnitude higher than the numerical diffusivity we estimated in \autoref{s:magfield_amp_largescale}. Similar order of magnitude estimates for the magnetic diffusivity are derived by \cite{2020arXiv200614607M}. Both these studies also find that Ohmic resistivity and the Hall effect are sub-dominant compared to ambipolar diffusion for the weakly-ionised primordial gas (see also, \citealt{2009ApJ...703.1096S}). Thus, ambipolar diffusion can decrease the strength of the large-scale dynamo by several orders of magnitude ($10^6$ in the outer disc, and $10^2$ in the inner disc; cf.~\autoref{eq:R_alphaRomega}), possibly even quenching it altogether when the field reaches $\sim1\,\rm{G}$.

However, in our strong field simulations where the field strength reaches $\sim1\,\rm{G}$, we did not see any amplification in any event, even with ideal MHD, simply because the field is already at or above saturation level when the disc forms. Thus, the relevant question is whether we would expect non-ideal effects to be significant for the field strengths typical of the weak-field cases. There, the field is $\sim\!10^{-3}\,\rm{G}$ when the sink particle forms (cf., top panel of \autoref{fig:dynamo}), and the disc that forms thereafter has characteristic densities $10^{10}$--$10^{13}\,\rm{cm^{-3}}$ (similar to what they are in the strong-field cases; cf., \autoref{fig:thermprofiles}). Since the effective resistivity provided by ambipolar diffusion scales as $\eta_{\mathrm{T}} \propto B$, for these physical conditions \citet{2019MNRAS.488.1846N} find resistivities of order $10^{18}$--$10^{20}\,\rm{cm^2\,s^{-1}}$, slightly smaller than our estimate of the numerical resistivity in the simulations. Thus, we conclude that non-ideal suppression of the dynamo is unimportant at least for the initial stages of amplification starting from a weak, sub-saturation field, consistent with the findings of \cite{2012ApJ...754...99S}.

A more subtle question is whether non-ideal effects might become important once significant amplification has taken place, and might thereby reduce the saturation field strength compared to what we find in our ideal MHD models. In the weakJ64 simulations where we see amplification, the runs saturate at peak field strengths of $\sim 0.1$ G, corresponding in \citeauthor{2019MNRAS.488.1846N}'s models to resisitivities of $\sim 10^{20}$--$10^{22}$ cm$^2$ s$^{-1}$ for the range of densities found in our discs. At the high end of the range, this is well in excess of our numerical resistivity, but seems unlikely to be sufficient to quench the dynamo: since $D_{\alpha\Omega}\propto \eta^{-2}_{\rm{T}}$, an increase from $\eta_{\mathrm{T}} \sim 10^{20}$ cm$^{2}$ s$^{-1}$ (our numerical value) to $\eta_{\rm{T}} \sim 10^{22}$ cm$^2$ s$^{-1}$ (from \citeauthor{2019MNRAS.488.1846N}'s models) at large disc radii, where the resistivity is largest, would correspond to a reduction in $D_{\alpha\Omega}$ by a factor of $\sim 10^4$ at those radii (cf.~\autoref{fig:dynamo_number}). While this represents a weakening of the dynamo, even the reduced figure is well above the threshold for efficient large-scale dynamo amplification. We therefore tentatively conclude that non-ideal effects are unlikely to substantially reduce the saturated field strength compared to that found in our ideal simulations.

Non-ideal MHD effects can also have consequences for the formation and evolution of accretion discs around the first stars, though there has been limited exploration of this effect in the literature. Nevertheless, we can gain insight by considering work on present-day (Pop I) star formation where non-ideal MHD effects have been explored in detail (\citealt{2018FrASS...5...39W} and \citealt{2020SSRv..216...43Z}, and references therein). Recent SPH simulations find that non-ideal MHD effects can be quite significant for the evolution of accretion discs around low-mass Pop~I stars \citep{2017MNRAS.466.1788W,2020MNRAS.495.3795W,2021arXiv210104129W}; in particular, these simulations find that larger and more massive discs form in the non-ideal MHD runs as compared to the ideal MHD runs when the turbulence is sub- or trans-sonic. Certain simulations also find that outflows are weaker in non-ideal MHD runs as compared to ideal MHD runs \citep[e.g.,][]{2016A&A...587A..32M,2019MNRAS.486.2587W,2020ApJ...900..180M,2021MNRAS.tmp..349X}. The inclusion of non-ideal effects also solves the magnetic-breaking catastrophe by suppressing the angular momentum transport that inhibits the formation of discs in ideal MHD \citep[e.g.,][]{2015ApJ...801..117T,2015MNRAS.452..278T,2016MNRAS.460.2050Z,2016A&A...587A..32M,2018A&A...615A...5V,2019MNRAS.489.1719W}. While ideal MHD does not hinder the formation of discs around first stars even in the case of strong fields (most likely due to the misalignment between the rotation axis and the tangled magnetic field in the presence of turbulence, as for Pop~I stars; e.g., \citealt{2013MNRAS.432.3320S,2015ApJ...810L..26T,2017PASJ...69...95T,2019MNRAS.489.1719W}), it is an open question how the properties of the discs might change when non-ideal effects are taken into account. We speculate that Pop III discs are closer to the case of discs around massive stars in the present-day Universe, because both types of systems are characterised by the overwhelming dominance of gravity over thermal pressure. This in turn makes the ram pressure of the accretion flow a dominant force that governs the physical properties of the disc \citep[e.g.,][]{2016MNRAS.463.2553R,2020AJ....160...78R}.

In summary, it is clearly desirable to run a large suite of non-ideal MHD simulations to study the formation of the first stars, but the associated computational cost (due to the combination of required high resolution, large statistics, and small timesteps due to magnetic diffusion) restricts us from performing such simulations at this time. We hope to remedy this in the future, and explore how dynamo amplification is impacted by non-ideal MHD effects during the formation of the first stars through 3D simulations.

\section{Implications for the IMF of the first stars}
\label{s:fragmentation}

\begin{figure*}
\includegraphics[width=\linewidth]{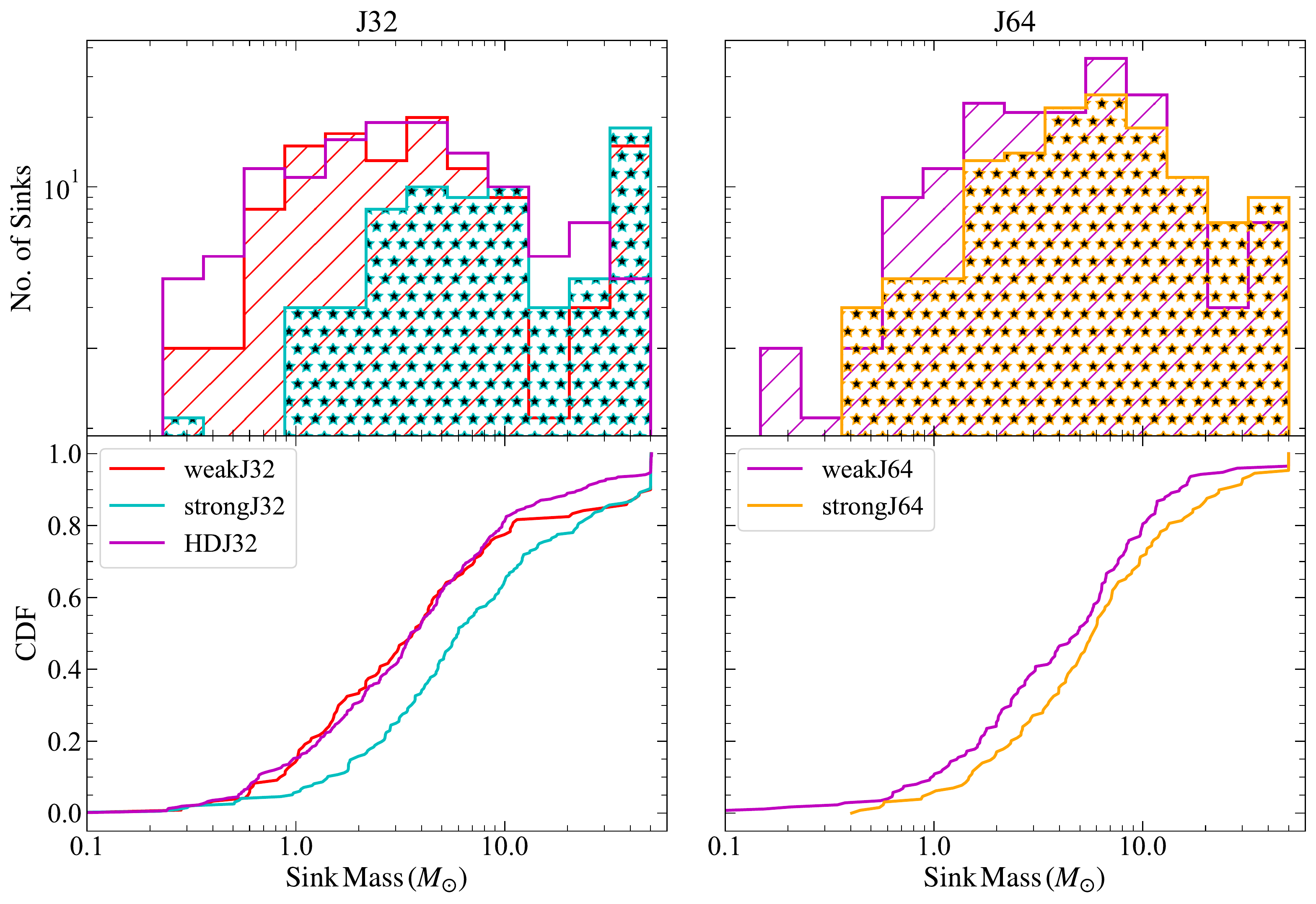}
\caption{\textit{Left panel}: The mass distribution (top) and the cumulative distribution (bottom) of sink particles that form till SFE  = 5 percent in 25 realizations in the weak- and strong-field runs with 32 cells per Jeans length. We also show the distribution for HDJ32 (without magnetic fields), adopted from \citetalias{2020arXiv200211502S}. \textit{Right panel}: the same distributions resulting from runs with 64 cells per Jeans length.}
\label{fig:sinkpdfcdf}
\end{figure*}

While the presence or absence of a dynamo in primordial accretion discs is interesting in itself, the main astrophysical question in which we are interested is how any resulting magnetic fields might affect the IMF of the first stars. This is something that does have at least potentially observable consequences. To investigate this question, we collect information on the sink mass distribution of all the four simulation categories: weakJ32, weakJ64, strongJ32, strongJ64, as well as the control case from \citetalias{2020arXiv200211502S}, which did not include a magnetic field and the Jeans length was resolved by 32 cells; we refer to this as the HDJ32 case. As opposed to \autoref{s:analysis_discussions} where we only used a subset of our simulations to study field amplification, we use all the simulations in each category to study the sink mass distribution and the primordial IMF.

The total number of sink particles (used as a proxy for stars) formed in weakJ32, weakJ64, strongJ32 and strongJ64, over the 25 realisations, are 121, 175, 70 and 130, respectively. This implies that higher Jeans resolution leads to more fragmentation in the MHD runs, by as high as a factor of 2. It is not easy to pin-point the cause of this finding, because the simulations are highly chaotic and non-linear. However, broadly speaking, we can attribute this effect to the fact that the accretion discs around the primary sink, and thus disc instabilities and sub-structure, are better resolved in J64 runs as compared to J32 runs. Given this result, we compare the sink mass distributions for the runs with 32 and 64 cells per Jeans length separately, so that we can disentangle the effects of magnetic fields and resolution. While this approach means that we are not necessarily capturing the true amount of fragmentation, since simulations are not fully converged, it does allow us to test with confidence how magnetic fields and dynamo amplification shift the IMF.

The left panel of \autoref{fig:sinkpdfcdf} shows the sink mass distribution for simulations with 32 cells per Jeans length. It is straightforward to see that the sink mass distribution of the strongJ32 runs is different from the other two, while the weakJ32 and HDJ32 runs are very similar, at least for $M\lesssim10\,\mathrm{M}_\odot$. To confirm this visual impression quantitatively, we apply the Kolmogorov-Smirnoff (KS) test for each pair of the runs shown in this panel. This test returns a $p$-value that describes the confidence level with which we can rule out the null hypothesis that the masses in each pair of runs were drawn from the same underlying distribution. Following \cite{2019MNRAS.490..513S} and \citetalias{2020arXiv200211502S}, we classify two distributions to be significantly different, if the $p$-value is $< 0.01$. The $p$-values for the pairs HDJ32$-$weakJ32, HDJ32$-$strongJ32 and weakJ32$-$strongJ32 come out to be 0.55, $5\times10^{-5}$ and $8\times10^{-4}$, respectively. Thus, the sink mass distribution produced by the strong magnetic field runs has a different origin than that produced by the weak field and HD runs. This finding is consistent with that of \citetalias{2020arXiv200211502S}. However, we note that the mass distributions for $M\gtrsim10\,\mathrm{M}_\odot$ are much more similar between weakJ32 and strongJ32, both showing a significantly higher number of massive stars than HDJ32.

The right panel of \autoref{fig:sinkpdfcdf} shows the same distributions for the runs with 64 cells per Jeans length. Visually, the weakJ64 and strongJ64 distributions are much closer to one another than are the weakJ32 and strongJ32 cases. The $p$-value for the pair weakJ64$-$strongJ64 is 0.12, implying no statistically significant difference in fragmentation between the weak- and the strong-field runs at higher Jeans resolution. This is not entirely unexpected, given that weakJ64 runs show significant field amplification. Thus, we find that first star cores with an initial field that falls below equipartition by a factor of $\sim 10^{7}$ produce an IMF that is significantly different from those that start near equipartition when we do not resolve dynamo amplification, but that this difference greatly diminishes, to the point of statistical undetectability, when we do capture dynamo growth. As further evidence of this effect, we note that, while we do not have a set of non-magnetic simulations at 64 cells per Jeans length to enable a direct comparison, the weakJ64 run shows \textit{less} fragmentation, and higher mean masses, than the HDJ32 case, despite having higher resolution, which tends to favour \textit{more} fragmentation. Thus, the effect of the dynamo-amplified magnetic field in suppressing fragmentation outweighs the effect of increasing the resolution.

It is interesting to consider the implications of these findings in the context of previous work on the mass of first stars formed in simulations. For example, based on radiation hydrodynamics simulations of first stars, \citet[Figure~9]{2014ApJ...792...32S} find that the mass spectrum peaks around $20\,\rm{M_{\odot}}$, and that most stars that form the earliest (prior to subsequent fragmentation) are more massive than $20\,\rm{M_{\odot}}$. While our conclusion of a top-heavy primordial IMF matches theirs, the exact location of the peak of the IMF is different. This is not entirely unexpected given the vast differences in physical (magnetic fields versus radiation feedback) and numerical schemes (initial conditions, AMR versus SPH, resolution, etc.) used in the two works. \cite{2014ApJ...781...60H} produce a suite of 100 first stars from their 2D radiation hydrodynamics simulations, finding stars as massive as $1000\,\rm{M_{\odot}}$. \cite{2015MNRAS.448..568H} perform cosmological simulations to study the primordial IMF, finding that it is top-heavy and the distribution is bimodal, peaking at $25\,\rm{M_{\odot}}$ and $250\,\rm{M_{\odot}}$, respectively. This bimodality is a result of the thermal evolution followed by accreting stars during the initial runaway collapse phase. It is difficult to directly compare the results of \cite{2014ApJ...781...60H,2015MNRAS.448..568H} with ours, given that we do not follow the protostellar accretion beyond $50\,\rm{M_{\odot}}$ in our 3D simulations without radiative feedback, whereas \cite{2015MNRAS.448..568H} perform 2D simulations without magnetic fields that do not capture the effects of fragmentation. The mass spectrum resulting from SPH-based 3D simulations of \cite{2013MNRAS.433.1094S} and \cite{2016MNRAS.462.1307S} is also top-heavy, but the distribution is relatively flat at early times and steepens later on, with several low-mass first stars forming in few thousand yr. Thus, while all simulations converge on the top-heavy nature of the primordial IMF, the intrinsic differences between them makes it difficult to compare them on a quantitative basis. This is further complicated by the fact that most simulations lack the necessary statistics to build up a statistically converged mass distribution. 

It is also worth comparing the primordial IMFs resulting from simulations discussed above to the constraints provided by observations of metal-poor stars that are believed to have been impacted by the first supernovae. \cite{2018ApJ...857...46I} find that the elemental abundance patterns measured in their compiled sample of extremely metal-poor stars (\,[Fe/H] < -3) are best described by $< 40\,\rm{M_{\odot}}$ Pop~III supernovae, with the majority of the patterns best fitted by a $25\,\rm{M_{\odot}}$ hypernova, confirming earlier results from \cite{2011A&A...527A..65H} and \cite{2014ApJ...792L..32I}. Similarly, \cite{2019MNRAS.488L.109N} find that the elemental abundance pattern in SMSS J160540.18-144323.1 $-$ the most iron-poor star known to-date (\,[Fe/H] = $-6.2 \pm 0.2$), is best described by a $10\,\rm{M_{\odot}}$ Pop~III supernova. \cite{2016ApJ...833...21P} find that the abundance pattern in several ultra metal-poor stars (\,[Fe/H] < -4) is best described by Pop~III supernovae with progenitor masses of $20-28\,\rm{M_{\odot}}$. On the other hand, the abundance pattern of another extremely metal-poor star is best fitted by a $40-60\,\rm{M_{\odot}}$ Pop~III supernova \citep{2015ApJ...806L..16B}. \cite{2017MNRAS.465..926D} find that observations of metal-poor stars are not in agreement with a flat mass distribution of the first stars with masses between $10-300\,\rm{M_{\odot}}$. Overall, these results indicate that at most a small fraction of first stars were more massive than $40\,\rm{M_{\odot}}$, which is not inconsistent with the predictions from at least some of the simulations we discuss above. However, both the observed and modeled abundance patterns are subject to uncertainties and free parameters that can significantly alter the resulting best-fits \citep[e.g.,][]{2008ApJ...684..588F,2009ApJ...693.1780J,2010ApJ...724..341H,2013ARA&A..51..457N,2016MNRAS.463.1518A,2017A&A...597A...6N,2020MNRAS.498.3703M}. Additionally, several works have remarked on the importance of considering multiple enrichment scenarios where yields from more than one Pop~III supernova enriched the stars in their surroundings \citep{2018MNRAS.478.1795H,2019MNRAS.482.1204H,2021MNRAS.500.5214W}, which further complicates the abundance pattern comparison. Nevertheless, these analyses remain one of the very few indirect ways through which we can put observational constraints on the primordial IMF.

In summary, we conclude that even if only a weak magnetic seed field was present in primordial clouds, it will be quickly driven to saturation by dynamo action and becomes dynamically important during Pop~III star formation. This is important because it means that (1) strong magnetic fields were likely present during Pop~III star formation, and (2) they had a significant impact on the primordial IMF.

\section{Conclusions}
\label{s:conclusions}
In this work, we study how magnetic fields can be amplified through a dynamo mechanism both on small and large scales in the accretion discs around Population III stars. There is a growing consensus that seeds of primordial magnetic fields, no matter how weak, were present in the early Universe \citep{2012SSRv..166...37W,2016RPPh...79g6901S,2016A&A...594A..19P}, and that they can be exponentially amplified during the collapse of minihaloes at $z\sim 20-30$ \citep{2012ApJ...745..154T}. Recent analysis has also shown that if dynamically strong magnetic fields were present during Population III star formation, they will significantly reduce fragmentation, thereby changing the IMF of the first stars \citep{2020arXiv200211502S}. However, previous work has left unresolved the question of how strong can magnetic fields grow during first star formation, and thus of how strong magnetic effects on the IMF are likely to be. This uncertainty is largely a function of numerical limitations: resolving the amplification of magnetic fields by dynamo action requires far higher resolution than is traditionally used in simulations of gravitational collapse and fragmentation.

To address this question, we perform a series of simulations in which we systematically vary the resolution (32 and 64 cells per Jeans length) and the initial strength of the turbulent magnetic field ($1\,\mathrm{fG}$ and $30\,\mathrm{\mu G}$, see \autoref{tab:tab1}). We use a subset of our simulations that form a single (massive) star to study the action of the dynamo in the protosellar accretion discs. The simulations with initially strong magnetic fields are a control case; they do not show any small-scale dynamo operation at either Jeans resolution, implying that the field is already saturated, as expected given our choice of initial field strength. By contrast, in the simulations where the initial magnetic field is weak, we find that the small-scale dynamo acts in accretion discs around the sink particles, amplifying the turbulent field strength such that, by the time a few percent of the initial cloud has accreted, the field in the disc reaches near saturation values similar to those in the runs where we start with the field already at saturation (see \autoref{fig:dynamo}). However, we also find that the timing and strength of field amplification is sensitive to resolution: simulations with 64 cells per Jeans length yield earlier and stronger field amplification than their lower-resolution counterparts. 

We also find a strong, large-scale mean toroidal component of the field in all the simulations (see \autoref{fig:Bprofiles}) together with a non-zero poloidal component, which is likely due to the operation of a large-scale $\alpha\Omega$-type dynamo. In this type of large-scale dynamo, the $\Omega$ effect winds up the field in the toroidal direction due to differential rotation (shear), and the $\alpha$ effect regenerates and maintains the poloidal field. \autoref{fig:dynamo_number} shows that the $\alpha\Omega$ dynamo acts efficiently in the outer disc, where we resolve the disc scale height with enough cells to capture its operation. Our findings are consistent with those of \citet{2011ApJ...731...62F}, who suggest that fully capturing a dynamo process likely requires resolution of $\sim$ 30 cells per Jeans length. Overall our results suggest a picture in which protostellar cores containing only seed fields with no organised structure and an energy density $\sim 7$ orders of magnitude below equipartition experience rapid growth of the field via both the small-scale dynamo, which increases the turbulent field strength to $\sim 1-10$ percent of equipartition, and the $\alpha\Omega$ dynamo, which moves a significant fraction of the energy stored in the disorganised, small-scale field into an organised, large-scale toroidal component. Although our simulations use ideal MHD, we also consider the likely effects of non-ideal processes in the context of recent work estimating the effective diffusivity due to Ohmic dissipation, the Hall effect, and ambipolar diffusion \citep{2019MNRAS.488.1846N, 2020arXiv200614607M}. We tentatively find that these effects should not significantly impede dynamo amplification, because the diffusivity depends on the field strength. Thus non-ideal effects are very small when the dynamo becomes to operate, and even once the field strength saturates, the diffusivity is small enough that the dynamo number remains $\gg 1$. Confirmation of these conclusions, however, will have to await full 3D non-ideal MHD simulations.

The development of magnetic fields at $1-10$ percent of equipartition even in protostellar cores that begin far below equipartition has profound implications for the IMF of the first stars. \cite{2020arXiv200211502S} show that the presence of an initial near-equipartition field strongly reduces the fragmentation of first star discs, leading to an IMF that is significantly more top-heavy, and deficient in stars with mass $\lesssim 1 \mathrm{M}_\odot$ that might survive to the present day. Our simulations here show that, thanks to dynamo action, this effect operates even in cores where the initial field is many orders of magnitude smaller, and that simulations can capture this effect, if they reach sufficient resolution. Hence, we propose that a scenario where magnetic fields remain weak throughout a Population III star formation episode is likely unphysical: magnetic field effects are always non-negligible.

A more speculative implication from this would be that Population III star formation might be subject to significant magnetic field-induced feedback effects like magnetic bubbles or jets (\citealt{2004ApJ...603..401T,2006ApJ...647L...1M,2014prpl.conf..173L,2014prpl.conf..451F,2018MNRAS.477..127D}; see, however, \citealt{2019MNRAS.485.5532G,2019MNRAS.486.3741H,2020arXiv200614607M}), and that it should be possible to detect these effects in simulations provided the innermost parts of the disc are sufficiently resolved. As the first massive stars explode, the first supernova explosions are likely to bring the magnetic fields into the interstellar medium, while also enriching it with metals \citep{2007ApJ...670....1G,2009ApJ...698..155S,2013MNRAS.430.2854M}. The metal enrichment is expected to lead to the formation of lower-mass stars due to cooling via metals and dust grains (e.g., \citealt{2003Natur.422..869S,2003Natur.425..812B,2005ApJ...626..627O}). For these Population II stars, the magnetic fields built up by dynamos around the first stars may become even more dynamically significant, and more important to limiting fragmentation \citep{2014MNRAS.440.1551L}, due to the diminished role of thermal pressure in gas subject to efficient cooling. The fields may also be further amplified in the haloes where this process takes place, via the same basic dynamo mechanisms we have explored here \citep{2013MNRAS.432..668L,2019MNRAS.487.4525G}. Self-consistent models of such environments should therefore always aim to incorporate the magnetic fields.

\section*{Acknowledgements}
We thank the anonymous referee for their feedback, which significantly improved this work. We also thank Chris McKee and Issac Shlosman for their useful comments on a preprint of this paper, and Amit Seta for insightful discussions on large-scale dynamos. PS is supported by the Australian Government Research Training Program (RTP) Scholarship. CF and MRK acknowledge funding provided by the Australian Research Council (ARC) through Discovery Projects DP170100603 (CF) and DP190101258 (MRK) and Future Fellowships FT180100495 (CF) and FT180100375 (MRK), and the Australia-Germany Joint Research Cooperation Scheme (UA-DAAD; both CF and MRK). MRK also acknowledges support from an Alexander von Humboldt Research Award. DRGS thanks for funding via Fondecyt regular (project code 1201280), Millenium Nucleus NCN19\_058 (TITANs, part of Agencia Nacional de Investigación y Desarrollo - ANID), and the Basal Centro de Excelencía en Astrofísica y Tecnologías Afines (CATA) grant PFB-06/2007. The simulations and data analyses presented in this work used high-performance computing resources provided by the Australian National Computational Infrastructure (NCI) through projects \texttt{ek9} (CF) and \texttt{jh2} (MRK) in the framework of the National Computational Merit Allocation Scheme and the Australian National University (ANU) Allocation Scheme, and as part of a contribution by NCI to the ARC Centre of Excellence for All Sky Astrophysics in 3 Dimensions (ASTRO 3D, CE170100013). Parts of this paper were written during the 2019 ASTRO 3D writing retreat. The software FLASH used in this work was developed in part by the Department of Energy (DOE) National Nuclear Security Administration Advanced Simulation and Computing Program- and DOE Office of Science Advanced Scientific Computing Research-supported Flash centre for Computational Science at the University of Chicago. Analysis was performed in \texttt{ipython} \citep{PER-GRA:2007,10.5555/1593511} and \texttt{Jupyter} packages using \texttt{yt} \citep{2011ApJS..192....9T}, \texttt{numpy} \citep{oliphant2006guide} and \texttt{scipy} \citep{2020SciPy-NMeth}; plots were created using \texttt{Matplotlib} \citep{Hunter:2007,thomas_a_caswell_2019_3264781}. The SAO/NASA Astrophysics Data System (ADS) is a digital library portal for researchers in astronomy and physics, operated by the Smithsonian Astrophysical Observatory (SAO) under a NASA grant. 

\section*{Data availability statement}
The simulation data underlying this article are available on reasonable request to the authors.

\bibliographystyle{mnras}
\bibliography{references} 

\appendix

\section{Effects of Jeans resolution on cooling}
\label{s:app_thermalbubble}
\begin{figure}
\includegraphics[width=\columnwidth]{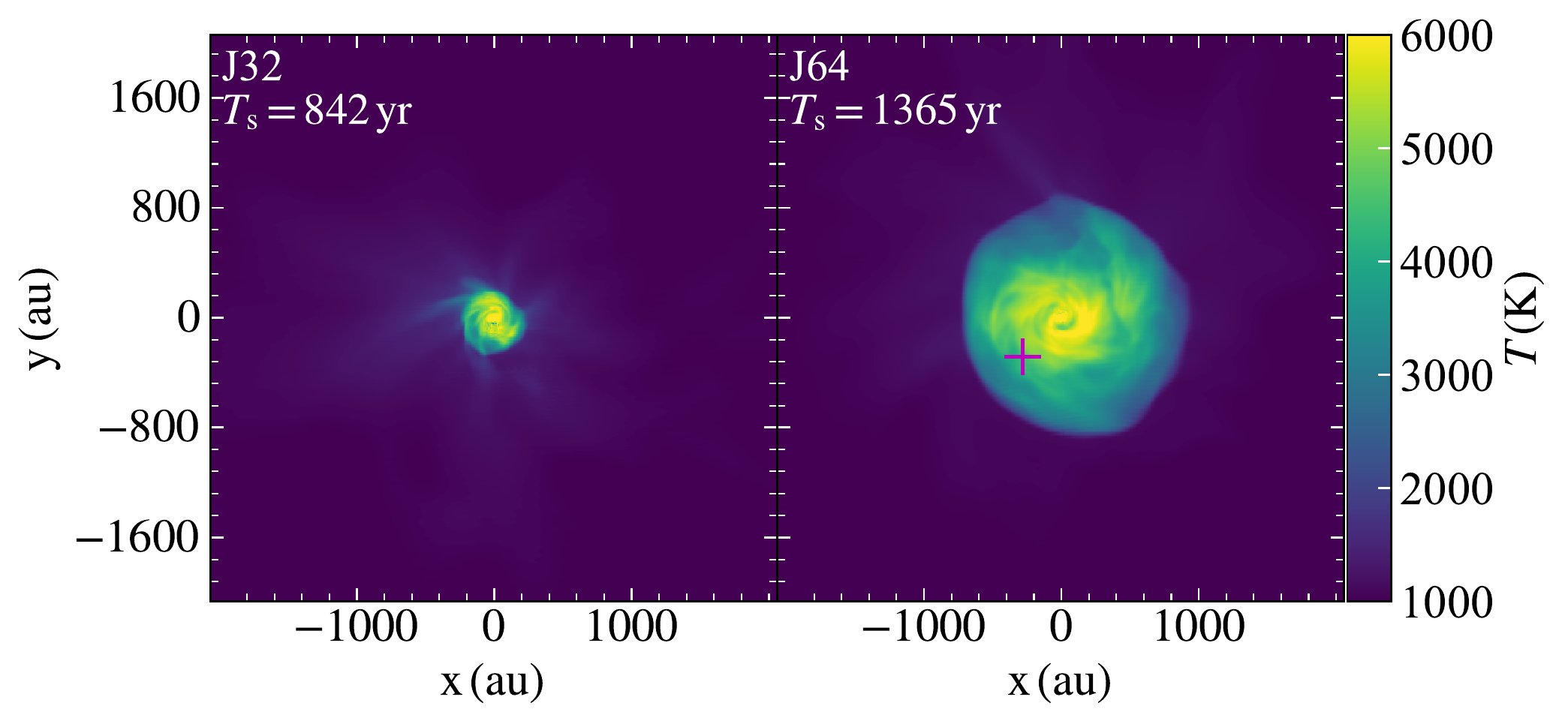}
\caption{Density-weighted projections of temperature for the J32 and J64 runs at the end of the simulation, when the SFE has reached 5 percent. The `+' marker denotes the sample point $p_1$ where we calculate the cooling length as the shock front travels through it earlier in the simulation.}
\label{fig:tbub_projections}
\end{figure}

The morphological evolution of the weak field runs changes significantly when we use 64 cells per Jeans length instead of 32. As discussed in \autoref{s:projections}, in the higher resolution case the simulation develops a  near-spherical bubble of gas at temperatures of $\approx 3000-6000$ K that expands over time; \citet{2012ApJ...745..154T} noticed a similar phenomenon in their highest-resolution simulations. To determine whether this bubble is associated with the presence of a magnetic field, we repeat the run shown in \autoref{fig:proj_dens} with identical gas initial conditions, but with no magnetic field, at resolutions of 32 (J32) and 64 (J64) cells per Jeans length. \autoref{fig:tbub_projections} shows the density-weighted temperature projections for the J32 and J64 runs. Given that we observe the same phenomenon as in the magnetic field runs, \text{i.e.,} a hot bubble appears in J64 but not in J32, we conclude that the presence of the bubble is not solely due to magnetic fields. 

\begin{figure}
\includegraphics[width=\linewidth]{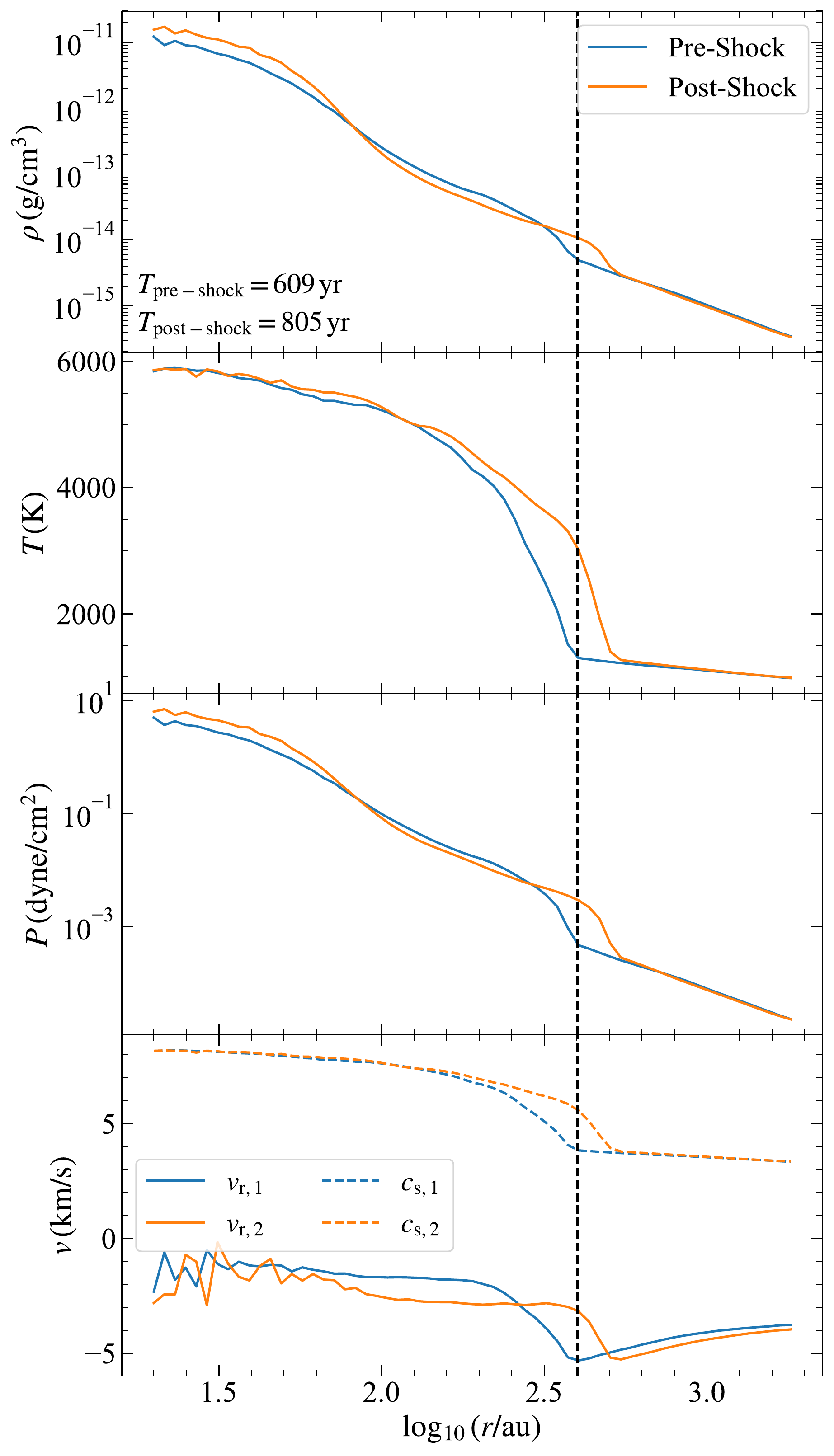}
\caption{Profiles of density, temperature, pressure and radial velocity along a radial ray passing through our sample point $p_1$ (\autoref{fig:tbub_projections}) at two times, just before (labelled ``Pre-Shock'') and just after (labelled ``Post-Shock'') the edge of the hot bubble reaches $p_1$, at a distance $r_1 = 400\,\mathrm{au}$ from the central star (indicated by the dashed vertical line). The time it takes for the gas to traverse the width of the shock is $204\,\mathrm{yr}$.}
\label{fig:tbub_radprofiles}
\end{figure}

Instead, we find that the key distinction between runs where we do and do not form bubbles is how well we resolve the temperature jump across the accretion shocks where matter falls onto the disc. To illustrate this point, we focus on a particular location inside the bubble, which we refer to as $p_1$ hereafter, at a radial distance of $r_1 = 400\,\mathrm{au}$ from the star, located in the plane of the disc, as indicated by the `+' in \autoref{fig:tbub_projections}. \autoref{fig:tbub_radprofiles} shows profiles of $\rho,\,T,\,P,\,v_{\mathrm{r}}$ and $c_{\mathrm{s}}$ along a radial ray passing through this point, at two times: just before and just after the bubble reaches $p_1$. We refer to the profile measured immediately before the bubble reaches our sample point as the ``Pre-Shock'' profile (blue in \autoref{fig:tbub_radprofiles}), and the one immediately after as the ``Post-Shock'' profile (orange in \autoref{fig:tbub_radprofiles}).

\begin{figure}
\includegraphics[width=\linewidth]{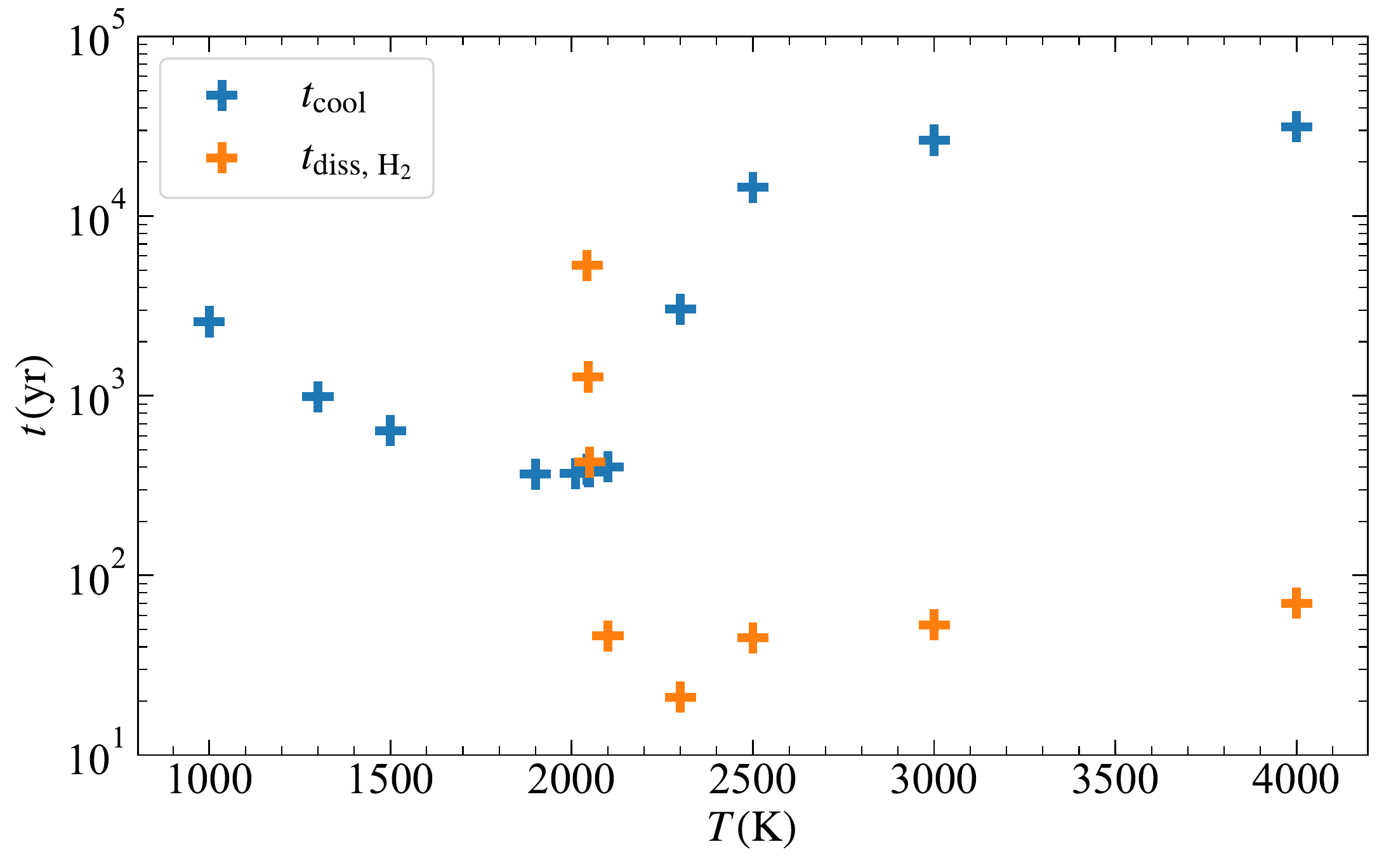}
\caption{The cooling and \hii dissociation timescales as a function of temperature, for the fixed post-shock chemical composition and density as listed in \autoref{tab:coolingrates}. At lower temperatures, $t_{\rm{diss\,H_2}}$ is infinity since there is no net dissociation of \hii. At higher temperatures, the molecular gas dissociates faster than it can cool.}
\label{fig:tbub_discool}
\end{figure}

\begin{table}
\caption{Pre-shock properties at point $p_1$ as obtained from \autoref{fig:tbub_radprofiles} in the J64 run. The quantity $x_q$ is the mass fraction of species $q$.}
\begin{tabular}{lcr}
\hline
Property & Pre-Shock & Post-Shock\\
\hline
$n\,\mathrm{(cm^{-3})}$ & $2.6\times10^9$&$6.2\times10^9$\\
$T\,\mathrm{(K)}$ & 1350 & 3110 \\
$x_{\mathrm{H}}$ & 0.76 & 0.76\\
$x_{\mathrm{H_2}}$ & $2.9\times10^{-3}$ & $1.2\times10^{-3}$\\
$x_{\mathrm{D}}$ & $4.6\times10^{-5}$ & $4.6\times10^{-5}$\\
$x_{\mathrm{HD}}$ & $4.0\times10^{-7}$ & $1.5\times10^{-7}$\\
$x_{\mathrm{H^+}}$ & $1\times10^{-8}$ & $4\times10^{-8}$\\
$x_{\mathrm{D^+}}$ & $4\times10^{-11}$ & $7\times10^{-11}$\\
\hline
$\Gamma_{\mathrm{rad}}\,\mathrm{(erg/cm^3/s)}$ & $4.8\times10^{-14}$ & $8.9\times10^{-17}$\\
$\Gamma_{\mathrm{chem}}\,\mathrm{(erg/cm^3/s)}$ & NA & $4.3\times10^{-15}$\\
$E_{\mathrm{T}}\,\mathrm{(erg/cm^3)}$ & $7.3\times10^{-4}$ & $3.8\times10^{-3}$\\
$t_{\mathrm{cool}}\,\mathrm{(yr)}$ & 477 & 27428\\
$t_{\mathrm{diss\,H_2}}\,\mathrm{(yr)}$ & $\infty$ & 55\\
\hline
\end{tabular}
\label{tab:coolingrates}
\end{table}

\autoref{tab:coolingrates} lists the properties of the gas at $p_1$ at times corresponding to the pre-shock and post-shock snapshots shown in \autoref{fig:tbub_radprofiles}. The ratios of densities, temperatures, and pressures in the pre- and post-shock conditions are as expected from the Rankine-Hugoniot jump conditions for non-radiative shocks. 

Using the post-shock values, we can calculate the total volumetric cooling rate via radiation, $\Gamma_{\mathrm{rad}}$, and via chemical reactions, $\Gamma_{\rm chem}$ (important at high temperature, where endothermic dissociation of H$_2$ is a significant coolant) from KROME. The time it will take for the gas to cool, $t_{\mathrm{cool}}$, depends on the cooling rate and the thermal energy per unit volume, $E_{\mathrm{T}}$,
\begin{equation}
t_{\mathrm{cool}} = \frac{E_{\mathrm{T}}}{\Gamma_{\mathrm{rad}} + \Gamma_{\rm chem}}\,,
\end{equation}
where  $E_{\mathrm{T}} = (3/2)nk_{\mathrm{B}}T$, and $k_{\mathrm{B}}$ is the Boltzmann constant\footnote{Note that the factor of $3/2$ implicitly assumes the gas is monoatomic, and thus ignores the effect of H$_2$ on the adiabatic index; given the very small H$_2$ fraction ($\sim 10^{-3}$) in the pre-shock gas, this approximation is reasonable.}. Similarly, the time it takes for \hii to dissociate can be given by,
\begin{equation}
t_{\mathrm{diss,\,H_2}} = \frac{x_{\rm{H_2}}}{-\dot x_{\rm{H_2}}}\,,
\end{equation}
where $x_{\rm H_2}$ is the H$_2$ mass fraction, and $\dot x_{\rm{H_2}}$ is the rate of change in the H$_2$ mass fraction; by convention, if $\dot{x}_{{\rm H}_2} \geq 0$, we take $t_{\rm diss,\,H_2} = \infty$. We see that the pre-shock conditions are characterised by rapid cooling ($t_{\rm cool} \sim 500$ yr) and no dissociation, while the post-shock conditions are characterised by much slower cooling ($t_{\rm cool} \sim 27,000$~yr) and rapid dissociation ($t_{\rm diss\,H_2} \sim 50$ yr). The reason for the much longer cooling time is the fact that, at the $\approx 3000$ K temperature found in the post-shock region, most collisions between H$_2$ molecules and H atoms lead to collisional dissociation rather than to excitation followed by radiative de-excitation.

In order to understand why resolution matters, it is helpful to consider how the cooling and dissociation times depend on temperature. \autoref{fig:tbub_discool} shows these quantities as a function of temperature for the post-shock chemical composition and density. The key feature to notice is that the thermal and chemical regime changes sharply at $\approx 2000$ K. Now, consider how material on the low-temperature side of this jump evolves as it encounters a shock. In the limit of infinite resolution, the shock has a width of the order of the particle mean free path. Given $n \sim 10^9\,\rm{cm^{-3}}$ and a typical cross-section for neutral species $\sim 10^{-16}\,\rm{cm^2}$, the shock width is $\sim 10^7\,\rm{cm}$. The time to traverse this distance at $\sim 1\,\rm{km\,s^{-1}}$ is $\sim 100\,\rm{s}$, which is tiny as compared to any radiative or chemical timescale. Thus, if this gas crosses a strong shock, its temperature increases by the usual factor $(\gamma+1)/(\gamma-1)$, without time for any radiative cooling to occur. If the gas is initially at $1300\,\rm{K}$, as is the case for our pre-shock sample point, this causes it to jump from the left to the right side of the $2000\,\rm{K}$ discontinuity in \autoref{fig:tbub_discool}. At that point, \hii dissociates faster than the gas is able to cool, and we get into the high-temperature, slow-cooling regime that characterises our post-shock region. Thus, the gas never cools.

Now, consider the case where the shock is broadened to a size $\sim 4\Delta x$, a typical shock width imposed by artificial viscosity (e.g., \citealt{2011MNRAS.415.3706C,2013MNRAS.432..711H}). If the resolution inside the region is $23\,\rm{au}$, as is the case in the J32 run, then the time required to traverse the shock region is greatly increased to $ \sim 92\,\rm{au} / 1\,\rm{(km\,s^{-1})} = 436\,\rm{yr}$. Interestingly, this is comparable to the pre-shock cooling time. The net effect is that the gas cools at the same time it is traversing the broadened shock, and thus never crosses over to the right side in \autoref{fig:tbub_discool}. It remains cool and with a significant fraction of H$_2$, exactly as we observe in the J32 run. On the other hand, if we double the Jeans resolution, then the time to traverse the shock is halved, and we are in the regime where the hydrodynamic time to cross the shock is smaller than the cooling time. Thus the temperature goes up, and we get to the right side of the jump at $2000\,\rm{K}$ in \autoref{fig:tbub_discool}, where $t_{\rm{diss\,H_2}} \ll t_{\rm{cool}}$. Once in this regime, the gas does not have enough time to cool before it dissociates, leading to the formation of a hot, H$_2$-poor bubble as we observe in the J64 run. This discussion also explains why a magnetic field, though not critical to the phenomenon we have identified, can nonetheless influence it: magnetic pressure helps mediate the shock (e.g., \citealt{2005ApJ...619..327F,2013ApJ...774..133L}), and thus changes the rate at which gas heats or cools as it passes the shock front. 

Thus, while our motivation to use a higher Jeans resolution was to better resolve the action of the small-scale dynamo, this result, along with earlier findings of \cite{2012ApJ...745..154T}, implies that a higher Jeans resolution is also critical for capturing the thermal and chemical changes that occurs across shocks.

\bsp	
\label{lastpage}
\end{document}